\documentclass[aps,prl,twocolumn,superscriptaddress]{revtex4-2}

\usepackage{amsmath,amssymb,amsthm,mathtools}
\usepackage{bm,upgreek} 
\usepackage{xcolor}
\usepackage{multirow}
\usepackage{graphicx}
\usepackage{epstopdf}
\usepackage[colorlinks,linkcolor=blue,anchorcolor=blue,citecolor=blue,urlcolor=blue]{hyperref}

\newcommand{\red}[1]{{\color{red}{#1}}}

\newcommand{\nn}{{\nonumber}}

\newcommand{\av}[1]{\left\langle #1 \right\rangle}
\newcommand{\md}{\mathrm{d}}

\newcommand{\w}{\omega}

\begin{document}
\title{Nonreciprocal Shapiro steps in Josephson diodes}
\author{Da Wang} \email{dawang@nju.edu.cn}
\author{Qiang-Hua Wang} 
\affiliation{National Laboratory of Solid State Microstructures $\&$ School of Physics, Nanjing University, Nanjing 210093, China}
\affiliation{Collaborative Innovation Center of Advanced Microstructures, Nanjing University, Nanjing 210093, China}
\author{Congjun Wu} \email{wucongjun@westlake.edu.cn}
\affiliation{New Cornerstone Science Laboratory, Department of Physics, School of Science, Westlake University, Hangzhou 310024, Zhejiang, China}
\affiliation{Institute for Theoretical Sciences, Westlake University, Hangzhou 310024, Zhejiang, China}
\affiliation{Key Laboratory for Quantum Materials of Zhejiang Province, School of Science, Westlake University, Hangzhou 310024, Zhejiang, China}
\affiliation{Institute of Natural Sciences, Westlake Institute for Advanced Study, Hangzhou 310024, Zhejiang, China}

\begin{abstract}
Various microscopic mechanisms have been proposed for the Josephson diode effect, which are unified phenomenologically within a nonreciprocal resistively and capacitively shunted junction model.
We further systematically investigate the dynamical response of these Josephson diodes to radio-frequency AC driving current, focusing in particular on the nonreciprocity of Shapiro steps.
We show that the Shapiro step structures encode the information about the underlying diode mechanisms in both overdamped and noised underdamped junctions.
In particular, we propose the zero-voltate step center as a fingerprint of different Josephson diode mechanisms in the overdamped regime, which is captured analytically within a harmonic approximation for small AC amplitudes.
Our results provides a route to experimentally distinguish different Josephson diode mechanisms, including those with and without time-reversal symmetry breaking.
\end{abstract}
\maketitle

\emph{Introduction.}
The semiconductor diode, a building block of modern microelectronics, exhibits highly asymmetric resistance under current reversion.
Recently, such a nonreciprocal phenomenon was generalized to superconductors, called Josephson diode effect (JDE) \cite{Hu_PRL_2007}, which has attracted significant attention due to its nonreciprocal supercurrent with unequal critical currents $I_{c\pm}$ along opposite directions \cite{Jiang_NP_2022,Nadeem_NRP_2023,Nagaosa_ARCMP_2024}.
JDE enables nonreciprocal supercurrent, offering potential applications in superconducting electronic circuits.
In the past few years, there have been many experiments observing the diode effect \cite{Ando_N_2020,Lyu_NC_2021,Turini_NL_2022,Strambini_NC_2022,Golod_NC_2022,Jeon_NM_2022,Narita_NN_2022,Lin_NP_2022,Pal_NP_2022,Wu_N_2022,Jiang_PRA_2022,Chiles_NL_2023,Gupta_NC_2023,DiezMerida_NC_2023,Gutfreund_NC_2023,Matsuo_NP_2023,Trahms_N_2023,Hou_PRL_2023,Zhao_S_2023,Li_AN_2024,Li_NC_2024,Reinhardt_NC_2024,Valentini_NC_2024,Anh_NC_2024,Ghosh_NM_2024,Le_N_2024,Wan_N_2024,Liu_SA_2024,Qi_NC_2025,Nagata_PRL_2025,Kudriashov_SA_2025,Shi_a_2025,Wang_NP_2025,Nagahama_PRL_2025,Zhang_PRX_2026,Chen_PRL_2026,Mangold_PRL_2026,Zhu_PRL_2026,Wei_NSR_2026}, which seems to be more universal than early proposals.
But on the theoretical side \cite{Misaki_PRB_2021,Scammell_2M_2022,Wang_a_2022,He_NJP_2022,Legg_PRB_2022,Fominov_PRB_2022,Kokkeler_PRB_2022,Daido_PRL_2022,Souto_PRL_2022,Zinkl_PRR_2022,Zhang_PRX_2022,Yuan_PotNAoS_2022,Davydova_SA_2022,He_NC_2023,Hu_PRL_2023,Steiner_PRL_2023,Lu_PRL_2023,Banerjee_PRB_2024,Banerjee_PRL_2024,Virtanen_PRL_2024,Mao_PRL_2024,Zeng_PRL_2025,Chakraborty_PRL_2025,Qi_PRB_2025}, a clear or full understanding of such an effect is still missing, partly because JDE is a property of asymmetry and thereby lacks a universal mechanism.

Nevertheless, it's widely accepted that JDE arises from breaking both inversion ($\+I$) and time-reversal ($\+T$) symmetries \cite{Jiang_NP_2022,Nadeem_NRP_2023,Nagaosa_ARCMP_2024}.
However, following this understanding, two questions can be raised immediately: (1) Are $\+I$- and $\+T$-breaking simultaneously sufficient to induce JDE? (2) Can diode effect occur if either $\+I$ or $\+T$ is preserving?
Regarding to the first question, the authors have shown that not only $\+I$ and $\+T$, but all the supercurrent reversion symmetries (such as particle-hole conjugation, time-reversal combined with spin or spatial rotations/reflections) have to be broken to realize JDE because any one of these symmetries would otherwise protect $I_{c+}=|I_{c-}|$ \cite{Wang_a_2022}.
Following this constraint, however, the answer to the second question seems to be negative.
But there indeed existed some experiments exhibiting the diode effect without applying magnetic field or proximate to ferromagnetism or breaking $\+T$ spontaneously, dubbed as the so-called ``zero-field'' diode effect \cite{Wu_N_2022,Li_AN_2024,Nagata_PRL_2025,Shi_a_2025,Zhang_PRX_2026,Zhu_PRL_2026}, which seems to be at odd with the above symmetry constraint.
One way to reconcile this paradox is that the experimental switching currents $I_{sw\pm}$ may not coincide with the critical current $I_{c\pm}$ but instead lie between $I_{c\pm}$ and retrapping currents $I_{r\pm}$ in the presence of inevitable noise, as proposed in $\+I$-breaking systems \cite{Misaki_PRB_2021}.
On the other hand, the current itself can also act as a source of $\+T$-breaking beyond linear response, e.g. through different boundary conditions between opposite current directions.
Such a JDE mechanism does not require intrinsic $\+T$-breaking of the junction itself, and has received attention only very recently \cite{Shi_a_2025,Zhang_PRX_2026,Nagata_PRL_2025}.
Given these complexities and subtleties, a unified framework to describe and distinguish various JDE mechanisms is highly desirable.

In this work, we describe different JDE mechanisms phenomenologically within a unified nonreciprocal resistively and capacitively shunted junction (RCSJ) model, and systematically investigate the dynamical response of these Josephson diodes to radio-frequency AC driving current. The nonreciprocal Shapiro steps have previously been studied in the overdamped limit within the RSJ (without capacitance) model with nonreciprocal current-phase relations \cite{Fominov_PRB_2022,Souto_PRL_2022,Hu_PRL_2023,Qi_PRB_2025}. Here, we consider and compare multiple JDE mechanisms within the generalized RCSJ framework, covering both overdamped and underdamped regimes. We find their Shapiro steps carry distinct signatures of the underlying mechanisms.


\emph{Generalized RCSJ Model.}
As a low-energy effective theory, various JDE mechanisms can be described by the generalized resistively and capacitively shunted junction (RCSJ) model
\begin{align}
C(\dot{\phi},\phi,I_{ext})\ddot{\phi} + \eta(\phi,I_{ext})\dot{\phi}+J_s(\phi,I_{ext})=I_{ext}
\end{align}
where $\phi$ is the Josephson phase, $C(\cdot)$ is the generalized capacitance, $\eta(\cdot)$ describes generalized damping, $J_s(\cdot)$ and $I_{ext}$ denote the Josephson and external current, respectively.
For simplicity, we set $\hbar/2e=1$ throughout this work.
Since there is only one degree of freedom $\phi$, it is enough to consider only two discrete symmetries here: inversion $\+I$ and time-reversal $\+T$. Under $\+I$, $\phi\to-\phi$ and $\dot{\phi}\to-\dot{\phi}$; under $\+T$, $\phi\to-\phi$ and $\dot{\phi}\to\dot{\phi}$. Since $I_{ext}$ is both $\+I$-odd and $\+T$-odd, we know that $C(\cdot)$ should be $\+I$-even and $\+T$-even, $J_s(\cdot)$ should be $\+I$-odd and $\+T$-odd, $\eta(\cdot)$ should be $\+I$-even and $\+T$-odd if both $\+I$ and $\+T$ are preserved. For a standard Josephson junction, $C(\dot{\phi},\phi,I_{ext})=C$, $\eta(\phi,I_{ext})=1/R$, and $J_s(\phi,I_{ext})=J_c\sin\phi$, for which the physics is controlled by the Stewart-McCumber parameter $\beta_c=CJ_cR^2$ \cite{Tinkham__1996}.

\begin{table}[b]
\caption{Four typical JDE mechanisms, symmetry breaking conditions and nonreciprocal behaviors, as described by the four terms in Eq.~\ref{eq:rcsj}.}
\label{table}
\begin{tabular}{|p{0.8cm}|p{3.8cm}|p{1.8cm}|p{0.8cm}|p{0.8cm}|}
\hline
RCSJ terms & Nonreciprocal \qquad \quad mechanisms & Broken \quad symmetries & $I_{c+}\ne |I_{c-}|$ & $I_{r+}\ne |I_{r-}|$ \\
\hline
$K$ & Asymmetric Josephson energy & $\+T$(intrinsic), $\+I$ & yes & yes \\
\hline
$b$ & External current effect on Josephson energy & $\+T$(extrinsic), $\+I$ & yes & yes \\
\hline
$a$ & Junction voltage effect on Josephson energy & $\+I$ & no & yes \\
\hline
$C_3$ & Asymmetric charge energy & $\+I$ & no & yes \\
\hline
\end{tabular}
\end{table}

Symmetry breaking allows more terms. If $\+I$ is broken, the mass term $C(\cdot)$ can contain $\+I$-odd terms such as $2C_3\dot{\phi}$ originating from asymmetric charge/kinetic energy $C_3\dot{\phi}^3/3$ \cite{Misaki_PRB_2021}. Meanwhile, the damping term $\eta(\cdot)$ can also contain $\+I$-odd terms such as $a\sin\phi$ originating from asymmetric voltage effect on the Josephson energy \cite{Hu_PRL_2007}.
If both $\+I$ and $\+T$ are broken, no constraint exists any more. A widely studied case is $K\cos2\phi$ in $J_s(\cdot)$ to account for asymmetric Josephson energy \cite{Fominov_PRB_2022,Souto_PRL_2022,Wang_a_2022}. Note that the harmonic term $\cos\phi$ can always be absorbed into the standard term $\sin\phi$ by a $\phi$-shift.
For small superconducting leads, particularly in some nanoscale experiments, the external current itself can effectively modify the Josephson energy \cite{Shi_a_2025,Nagata_PRL_2025}. Such an effect can be captured effectively by adding another term $bI_{ext}\sin\phi$ in the Josephson current $J_s(\cdot)$. Note that such a feedback term breaks $\+T$ globally including the whole circuit, but preserves $\+T$ of the junction itself.
In the following, we will focus on the above four mechanisms of JDE ($C_3$, $a$, $b$, $K$), as summarized in table~\ref{table}. Putting these terms together in the equation of motion, we have
\begin{align}
&(3C_4\dot{\phi}^2+2C_3\dot{\phi}+C)\ddot{\phi}+\dot{\phi}/R \nn\\
&\quad\quad+(J_c+a\dot{\phi}+b I_{ext})\sin\phi-K\cos2\phi = I_{ext}
\label{eq:rcsj}
\end{align}
where we have also added a nonlinear capacitance term $\frac14C_4\dot{\phi}^4$ in the charge energy to ensure its positivity.
In addition, to avoid unphysical result, we always require the total Josephson coupling $J_{eff}=J_c+a\dot{\phi}+bI_{ext}$ within the range from $0$ to $2J_c$.
In this work, the external current is generally set as $I_{\rm ext}(t)=I_{\rm dc}+I_{\rm ac}\sin(\w t)+I_{\rm noise}(t)$ where the noise current satisfies $\av{I_{\rm noise}(t)I_{\rm noise}(t')}=D\delta(t-t')$.

\emph{Two types of JDE.}
Let us first consider the DC effect by setting $I_{ext}=I_{dc}$.
According to the literature, an ideal Josephson diode is defined by nonreciprocal critical currents $I_{c+}\ne|I_{c-}|$. Setting $\dot{\phi}$=0, we have
$I_{dc}=(J_c\sin\phi - K\cos2\phi)/(1-b\sin\phi)$.
The critical currents $I_{c\pm}$ are two bounds of $I_{dc}$ by varying $\phi$, i.e. $I_{c+}=\text{max}(I_{dc})$ and $I_{c-}=\text{min}(I_{dc})$.
Clearly, both nonzero $K$ and $b$ can give the ideal JDE.
There is an intensive debate in the community that whether $\+T$-breaking is necessary for an ideal JDE?
Our answer is always yes if we consider the whole circuit, but depends on the mechanism if we consider the junction itself (yes for $K\ne0$ and no for $b\ne0$).

Another type of JDE is defined by nonreciprocal retrapping current $I_{r+}\ne|I_{r-}|$ even if $I_{c+}=|I_{c-}|$. Both $a\ne0$ and $C_3\ne0$ belong to this type. Their effect on $I_{r\pm}$ can be understood in the underdamped limit $\beta_c\to\infty$, in which case $I_{r}\approx4J_c/\pi\sqrt{\beta_c}$ for a standard RCSJ model \cite{Tinkham__1996}. For $a\ne0$, the effective damping $R^{-1}+a\sin\phi$ becomes direction dependent; while for $C_3\ne0$, the effective capacitance $C+2C_3I_{r\pm}R$ is nonreciprocal. In real experiments, noise current is always inevitable and will cause randomly tunneling between the zero-voltage and finite-voltage metastable states, resulting in nonreciprocal switching currents $I_{sw+}\ne|I_{sw-}|$.

\begin{figure}
\includegraphics[width=\linewidth]{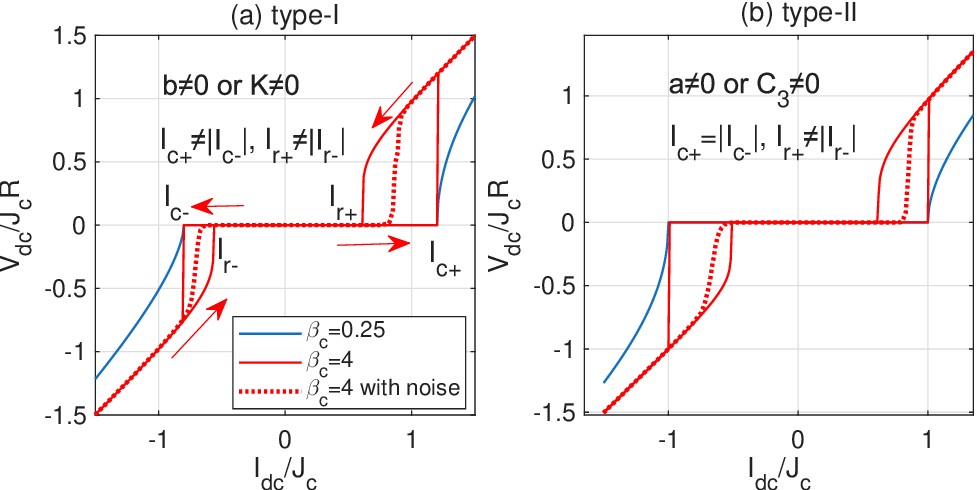}
\caption{Two types of DC Josephson diodes for overdamped ($\beta_c=0.25$) and underdamped ($\beta_c=4$) junctions. For the latter, noise is added to eliminate the hysteresis behavior and gives rise to nonreciprocal switching current for all the cases.}
\label{fig:DC}
\end{figure}

In Fig.~\ref{fig:DC}, we show the two types of JDE in both overdamped ($\beta_c=0.25$) and underdamped ($\beta_c=4$ with hysteresis) regimes. For the latter, diode behavior is clearly seen after noise average for both types of junctions. Therefore, it is difficult to distinguish these various JDE mechanisms only by DC experiments. For this reason, we move on to investigate their dynamical response to AC drivings.

\begin{figure*}
\includegraphics[width=\textwidth,trim={60 10 60 10},clip]{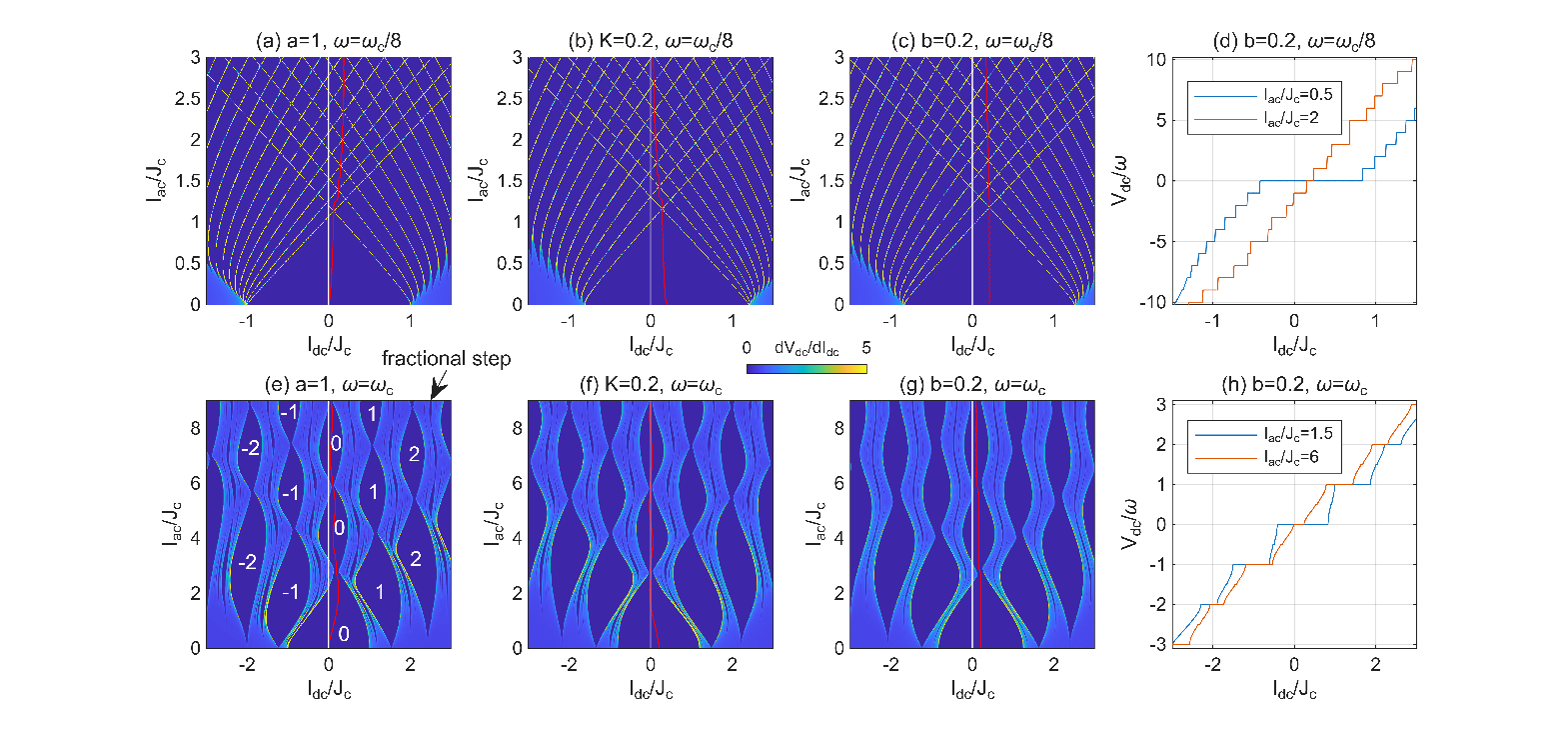}
\caption{Shapiro steps in overdamped Josephson diodes with $\beta_c=0.25$. (a-c) show $dV_{dc}/dI_{dc}$ on the $(I_{dc},I_{ac})$ plane for nonzero $a$, $K$, and $b$, respectively, at low frequency $\w=\w_c/8$. The zero-voltage step centers are plotted by the red lines. In (d), two typical $V_{dc}-I_{dc}$ curves are shown, displaying shift Shapiro steps along the current direction. (e-h) are similar to (a-d) but for a high frequency $\w=\w_c$.}
\label{fig:od}
\end{figure*}

\emph{Shapiro steps in overdamped junctions.}
We solve the equation of motion Eq.~\ref{eq:rcsj} numerically by setting the external current $I_{ext}(t)=I_{dc}+I_{ac}\sin(\w t)$ to obtain $\phi(t)$ and $V(t)$ from which $V_{dc}=\av{V(t)}$ is extracted. At low frequency $\w=\w_c/8$ ($\w_c=JR$ as the characteristic frequency of the junction), $dV_{dc}/dI_{dc}$ are plotted on the $(I_{dc},I_{ac})$ plane in Fig.~\ref{fig:DC}(a-c) for nonzero $a$, $K$ and $b$, respectively. The result of $C_3\ne0$ is not shown directly since the capacitance plays less role than the other mechanisms.
Different Shapiro steps with $V_{dc}=n\w$ are separated by a series of lines. As an example for clarity, two typical $V_{dc}-I_{dc}$ curves are plotted in Fig.~\ref{fig:DC}(d) for the case of nonzero $b$. Clearly, nonreciprocity does not change the step voltage quantization, since the running frequency is still locked at $n\w$. Nevertheless, the steps are shifted along $I_{dc}$. In particular, the zero-voltage step can shift completely across the origin, leading to ``perfect diode effect'' if we follow the usual definition of diode coefficient $Q=(|I_{c+}|-|I_{c-}|)/(|I_{c+}|+|I_{c-}|)$. Furthermore, we find the zero step center, denoted by $I_0$, as indicated by red lines, exhibits quite different behaviors for the three cases: with increasing the AC amplitude $I_{ac}$, $|I_0|$ increases for $a\ne0$, decreases for $K\ne0$, while keeps almost unchanged for $b\ne0$. These differences can be qualitatively understood as follows.
For $a\ne0$, the nonreciprocity comes from direction-dependent effective damping $\eta=R^{-1}+a\sin\phi$ and is only activated by phase motion.
For $K\ne0$, the nonreciprocity originates from the asymmetric Josephson energy, hence, stronger $I_{ac}$ drives $\phi$ to visit a broader range and thus reduces the net nonreciprocity.
For $b\ne0$, the nonreciprocal term $bI_{ext}\sin\phi$ grows with $I_{ac}$, hence, compensating the increased phase excursion and resulting in an almost unchanged $I_0$ as observed in our numerical results.

At high frequency $\w=\w_c$, the results are shown in Fig.~\ref{fig:od}(e-h). (Much higher frequencies are not considered here, as the Josephson coupling is expected to be significantly suppressed.) The Shapiro steps are still quantized at $V_{dc}=n\w$, while are now separated by continuum (not quantized) regimes, as clearly seen in the $V_{dc}-I_{dc}$ curves in Fig.~\ref{fig:od}(h). The steps are also shifted along $I_{dc}$, and the tendencies of zero-voltage step center $I_0$ (see red lines) with $I_{ac}$ exhibit similar behaviors to the above low frequency results. In addition, there are fractional steps within the continuum regimes, including the prominent half-integer steps and some other rational ones such as $1/3$. Actually, the fractional steps also exist in the standard RCSJ model, and therefore should not by themselves be taken as evidence for a higher-harmonic current-phase relation.

\begin{figure*}
\includegraphics[width=0.9\textwidth]{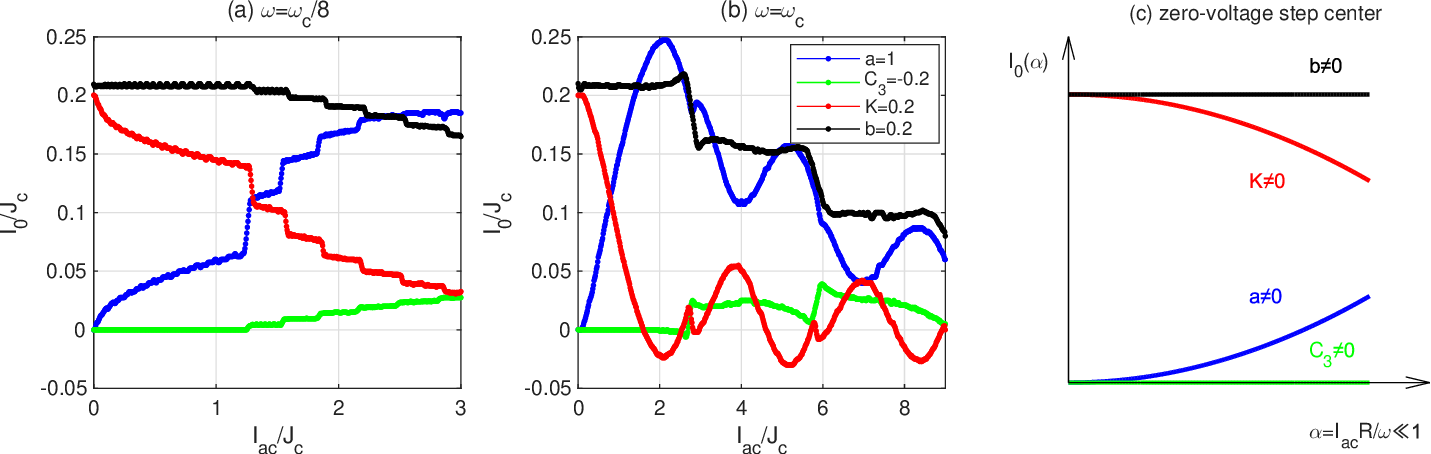}
\caption{Zero-voltage step center $I_0$ as a fingerprint of various overdamped Josephson diodes. From the RCSJ simulations in Fig.~\ref{fig:od}, we extract $I_0$ versus $I_{ac}$ as plotted in (a) for $\w=\w_c/8$ and in (b) for $\w=\w_c$, respectively. At small $I_{ac}$, these results can be captured by the harmonic approximation as sketched in (c).}
\label{fig:I0}
\end{figure*}

\emph{Harmonic approximation.}
We have shown that the zero-voltage step center $I_0$ can be taken as a fingerprint to distinguish the three JDE mechanisms ($a$, $K$, $b$) for overdamped junctions. In Fig.~\ref{fig:I0} (a) and (b), we directly plot $I_0$ versus $I_{ac}$ for $\w=\w_c/8$ and $\w_c$, respectively. Here, we additionally add the results of $I_0$ for $C_3\ne0$, which keep at zero for small $I_{ac}$, and only grows up for large $I_{ac}$. This can be roughly understood by the overdamped condition under which the strong damping dominates over the weak inertia (capacitance).

In the overdamped limit, the capacitance terms can be neglected, resulting in the so-called RSJ model, for which we can look for locked-phase solutions under the harmonic approximation \cite{Tinkham__1996}
\begin{align}
\phi(t)=n\w t + \delta - \alpha\cos\w t,
\end{align}
which gives voltage quantization $V_{dc}=\langle\dot{\phi}\rangle=n\w$.
After some straightforward algebraic calculations by taking $a$, $b$ and $K$ as perturbations, we obtain the zero-voltage step center
\begin{align}
I_0=\frac{a}{2}\alpha_0^2RJ_c+bJ_c+K(1-\alpha_0^2) + \+O(\alpha_0^4)
\end{align}
where $\alpha_0=I_{ac}R/\w$. Clearly, such a harmonic approximation successfully captures the main features of the four mechanisms, as sketched in Fig.~\ref{fig:I0}(c).
More details about the justifications and calculations of the harmonic approximation can be found in the Supplementary Material.

\begin{figure*}
\includegraphics[width=0.9\textwidth,trim={60 20 60 20},clip]{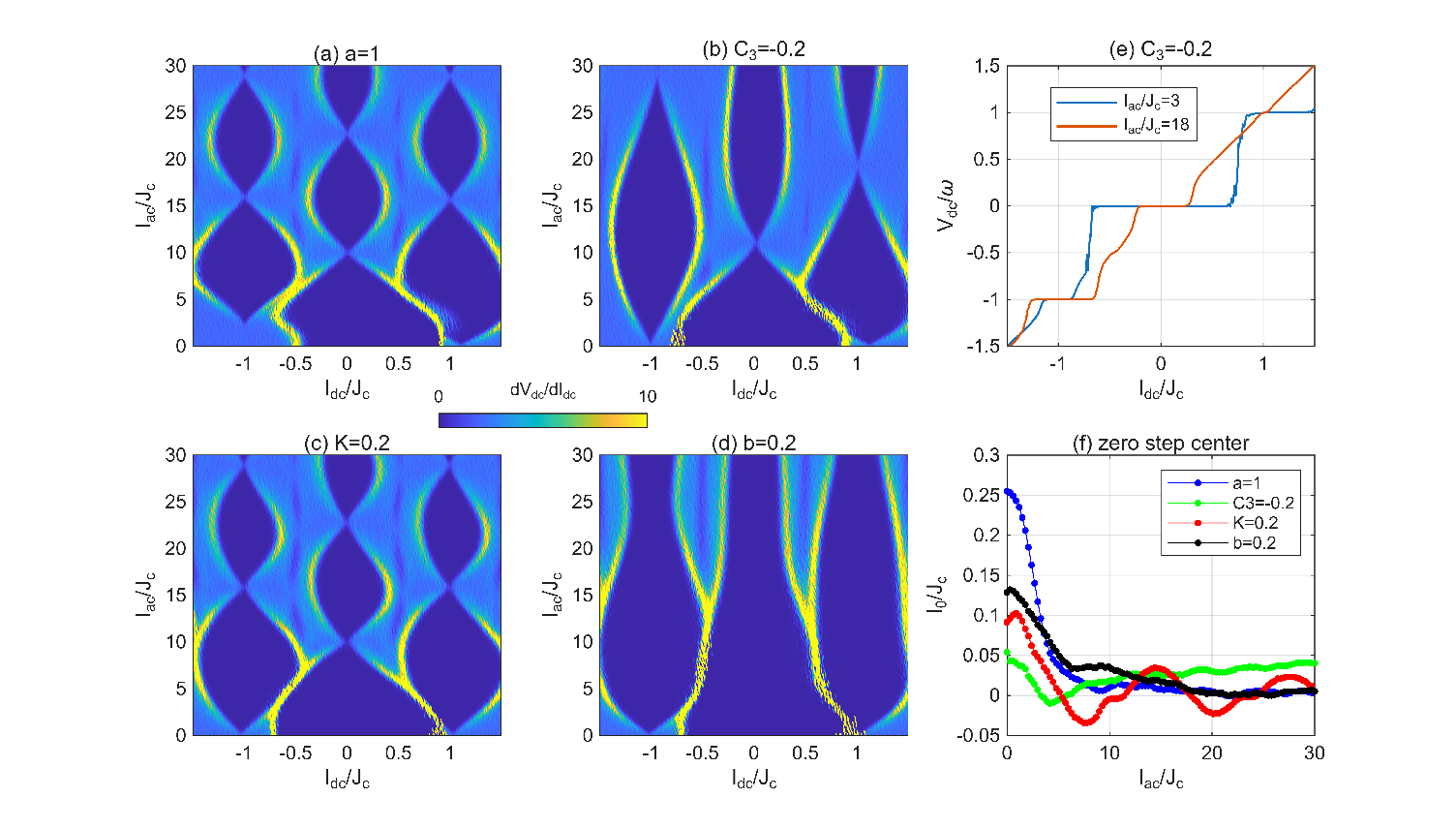}
\caption{Shapiro step structures in underdamped Josephson diodes under weak noise with $\beta_c=4$ at $\w=\w_c$. (a-d) show $dV_{dc}/dI_{dc}$ on the $(I_{dc},I_{ac})$ plane for nonzero $a$, $C_3$, $K$, and $b$, respectively. Two typical $V_{dc}-I_{dc}$ curves for $C_3=-0.2$ are displayed in (e). The zero-voltage step center $I_0$ in (a-d) are extracted and plotted versus $I_{ac}$ in (f).}
\label{fig:ud}
\end{figure*}

\emph{Underdamped junctions.}
For underdamped junctions, the phase inertia (capacitance) becomes more important, leading to multi-stability and more complex dynamical responses. For low-frequency AC drives, multiple step solutions, including both integer and non-integer, are found to coexist and noise can easily destroy the sharp stair-like structures in the $V_{dc}-I_{dc}$ curves as in the overdamped regime. In contrast, for high-frequency AC drives such as $\w=\w_c$, stable integer-valued Shapiro steps are found to persist despite the strong nonlinear dynamics.

In Fig.~\ref{fig:ud}(a-d), we plot $dV_{dc}/dI_{dc}$ on the $(I_{dc},I_{ac})$ plane in the presence of weak noise, with nonzero $a$, $C_3$, $K$ and $b$, respectively. The Shapiro steps can be readily identified although the step edges become slightly blurred by the noise.
Different from the overdamped case, all these four mechanisms exhibit nonreciprocal switching currents in the limit of $I_{ac}\to0$. Therefore, their distinct signatures only emerge at finite $I_{ac}$.
The $C_3\ne0$ case can be easily identified by its obvious asymmetry between $n$ and $-n$ step regions. In particular, as shown in Fig.~\ref{fig:ud}(e), the lengths of $\pm1$ steps differ significantly. Such a strong nonreciprocity can be understood in terms of the voltage-asymmetric effective capacitance $C_{eff}=C+2C_3V_{dc}$.
The $b\ne0$ case also exhibits a distinct feature that regions with the same $n$ become connected together due to the strong modification to the conventional harmonic Josephson coupling.
The cases with $a\ne0$ and $K\ne0$ do not exhibit fundamentally distinct features, but they can also be distinguished by extracting the zero-voltage step center $I_0$ as shown in Fig.~\ref{fig:ud}(f).
For $K\ne0$, $I_0$ oscillates around zero with increasing $I_{ac}$, while for $a\ne0$, $I_0$ gradually drops to zero without sign change.

\emph{Summary and discussions.}
In this work, we describe different JDE mechanisms within a nonreciprocal RCSJ model.
By systematically investigating their dynamical response to AC drivings, we find the Shapiro step structures carry distinct signatures of the underlying mechanisms.
In particular, we propose the zero-voltage step center as a fingerprint of the underlying diode mechanism, which can be captured by the harmonic approximation for small AC amplitudes.

Before closing, we would like to give some remarks about the symmetries. In the RCSJ model with only one degree of freedom $\phi$, it is enough to consider only two symmetries, $\+I$ and $\+T$. For microscopic models, these two symmetries should be extended to two classes: $\+I$-like (unitary) and $\+T$-like (antiunitary). Therefore, breaking both $\+I$ and $\+T$ is only a necessary but not sufficient condition for the ideal diode effect with $I_{c+}\ne|I_{c-}|$. Similarly, breaking $\+I$ is only a necessary but not sufficient condition for an experimental diode effect with $I_{sw+}\ne|I_{sw-}|$.

\emph{Acknowledgments.}
This work is supported by National Key R\&D Program of China (Grant No. 2022YFA1403201, No. 2024YFA1408100) and National Natural Science Foundation of China (Grant No. 12234016, No. 12274205, No. 12374147, No. 92365203, No. 12174317).
This work has also been supported by the New Cornerstone Science Foundation.

\nocite{*}
\bibliography{diode}

\begin{thebibliography}{68}%
\makeatletter
\providecommand \@ifxundefined [1]{%
 \@ifx{#1\undefined}
}%
\providecommand \@ifnum [1]{%
 \ifnum #1\expandafter \@firstoftwo
 \else \expandafter \@secondoftwo
 \fi
}%
\providecommand \@ifx [1]{%
 \ifx #1\expandafter \@firstoftwo
 \else \expandafter \@secondoftwo
 \fi
}%
\providecommand \natexlab [1]{#1}%
\providecommand \enquote  [1]{``#1''}%
\providecommand \bibnamefont  [1]{#1}%
\providecommand \bibfnamefont [1]{#1}%
\providecommand \citenamefont [1]{#1}%
\providecommand \href@noop [0]{\@secondoftwo}%
\providecommand \href [0]{\begingroup \@sanitize@url \@href}%
\providecommand \@href[1]{\@@startlink{#1}\@@href}%
\providecommand \@@href[1]{\endgroup#1\@@endlink}%
\providecommand \@sanitize@url [0]{\catcode `\\12\catcode `\$12\catcode
  `\&12\catcode `\#12\catcode `\^12\catcode `\_12\catcode `\%12\relax}%
\providecommand \@@startlink[1]{}%
\providecommand \@@endlink[0]{}%
\providecommand \url  [0]{\begingroup\@sanitize@url \@url }%
\providecommand \@url [1]{\endgroup\@href {#1}{\urlprefix }}%
\providecommand \urlprefix  [0]{URL }%
\providecommand \Eprint [0]{\href }%
\providecommand \doibase [0]{https://doi.org/}%
\providecommand \selectlanguage [0]{\@gobble}%
\providecommand \bibinfo  [0]{\@secondoftwo}%
\providecommand \bibfield  [0]{\@secondoftwo}%
\providecommand \translation [1]{[#1]}%
\providecommand \BibitemOpen [0]{}%
\providecommand \bibitemStop [0]{}%
\providecommand \bibitemNoStop [0]{.\EOS\space}%
\providecommand \EOS [0]{\spacefactor3000\relax}%
\providecommand \BibitemShut  [1]{\csname bibitem#1\endcsname}%
\let\auto@bib@innerbib\@empty
\bibitem [{\citenamefont {Hu}\ \emph {et~al.}(2007)\citenamefont {Hu},
  \citenamefont {Wu},\ and\ \citenamefont {Dai}}]{Hu_PRL_2007}%
  \BibitemOpen
  \bibfield  {author} {\bibinfo {author} {\bibfnamefont {J.}~\bibnamefont
  {Hu}}, \bibinfo {author} {\bibfnamefont {C.}~\bibnamefont {Wu}},\ and\
  \bibinfo {author} {\bibfnamefont {X.}~\bibnamefont {Dai}},\ }\bibfield
  {title} {\bibinfo {title} {Proposed {{Design}} of a {{Josephson Diode}}},\
  }\href {https://doi.org/10.1103/PhysRevLett.99.067004} {\bibfield  {journal}
  {\bibinfo  {journal} {Phys. Rev. Lett.}\ }\textbf {\bibinfo {volume} {99}},\
  \bibinfo {pages} {067004} (\bibinfo {year} {2007})}\BibitemShut {NoStop}%
\bibitem [{\citenamefont {Jiang}\ and\ \citenamefont
  {Hu}(2022)}]{Jiang_NP_2022}%
  \BibitemOpen
  \bibfield  {author} {\bibinfo {author} {\bibfnamefont {K.}~\bibnamefont
  {Jiang}}\ and\ \bibinfo {author} {\bibfnamefont {J.}~\bibnamefont {Hu}},\
  }\bibfield  {title} {\bibinfo {title} {Superconducting diode effects},\
  }\href {https://doi.org/10.1038/s41567-022-01701-0} {\bibfield  {journal}
  {\bibinfo  {journal} {Nat. Phys.}\ }\textbf {\bibinfo {volume} {18}},\
  \bibinfo {pages} {1145} (\bibinfo {year} {2022})}\BibitemShut {NoStop}%
\bibitem [{\citenamefont {Nadeem}\ \emph {et~al.}(2023)\citenamefont {Nadeem},
  \citenamefont {Fuhrer},\ and\ \citenamefont {Wang}}]{Nadeem_NRP_2023}%
  \BibitemOpen
  \bibfield  {author} {\bibinfo {author} {\bibfnamefont {M.}~\bibnamefont
  {Nadeem}}, \bibinfo {author} {\bibfnamefont {M.~S.}\ \bibnamefont {Fuhrer}},\
  and\ \bibinfo {author} {\bibfnamefont {X.}~\bibnamefont {Wang}},\ }\bibfield
  {title} {\bibinfo {title} {The superconducting diode effect},\ }\href
  {https://doi.org/10.1038/s42254-023-00632-w} {\bibfield  {journal} {\bibinfo
  {journal} {Nat. Rev. Phys.}\ }\textbf {\bibinfo {volume} {5}},\ \bibinfo
  {pages} {558} (\bibinfo {year} {2023})}\BibitemShut {NoStop}%
\bibitem [{\citenamefont {Nagaosa}\ and\ \citenamefont
  {Yanase}(2024)}]{Nagaosa_ARCMP_2024}%
  \BibitemOpen
  \bibfield  {author} {\bibinfo {author} {\bibfnamefont {N.}~\bibnamefont
  {Nagaosa}}\ and\ \bibinfo {author} {\bibfnamefont {Y.}~\bibnamefont
  {Yanase}},\ }\bibfield  {title} {\bibinfo {title} {Nonreciprocal {Transport}
  and {Optical} {Phenomena} in {Quantum} {Materials}},\ }\href
  {https://doi.org/10.1146/annurev-conmatphys-032822-033734} {\bibfield
  {journal} {\bibinfo  {journal} {Annu. Rev. Conden. Ma. P.}\ }\textbf
  {\bibinfo {volume} {15}},\ \bibinfo {pages} {63} (\bibinfo {year}
  {2024})}\BibitemShut {NoStop}%
\bibitem [{\citenamefont {Ando}\ \emph {et~al.}(2020)\citenamefont {Ando},
  \citenamefont {Miyasaka}, \citenamefont {Li}, \citenamefont {Ishizuka},
  \citenamefont {Arakawa}, \citenamefont {Shiota}, \citenamefont {Moriyama},
  \citenamefont {Yanase},\ and\ \citenamefont {Ono}}]{Ando_N_2020}%
  \BibitemOpen
  \bibfield  {author} {\bibinfo {author} {\bibfnamefont {F.}~\bibnamefont
  {Ando}}, \bibinfo {author} {\bibfnamefont {Y.}~\bibnamefont {Miyasaka}},
  \bibinfo {author} {\bibfnamefont {T.}~\bibnamefont {Li}}, \bibinfo {author}
  {\bibfnamefont {J.}~\bibnamefont {Ishizuka}}, \bibinfo {author}
  {\bibfnamefont {T.}~\bibnamefont {Arakawa}}, \bibinfo {author} {\bibfnamefont
  {Y.}~\bibnamefont {Shiota}}, \bibinfo {author} {\bibfnamefont
  {T.}~\bibnamefont {Moriyama}}, \bibinfo {author} {\bibfnamefont
  {Y.}~\bibnamefont {Yanase}},\ and\ \bibinfo {author} {\bibfnamefont
  {T.}~\bibnamefont {Ono}},\ }\bibfield  {title} {\bibinfo {title} {Observation
  of superconducting diode effect},\ }\href
  {https://doi.org/10.1038/s41586-020-2590-4} {\bibfield  {journal} {\bibinfo
  {journal} {Nature}\ }\textbf {\bibinfo {volume} {584}},\ \bibinfo {pages}
  {373} (\bibinfo {year} {2020})}\BibitemShut {NoStop}%
\bibitem [{\citenamefont {Lyu}\ \emph {et~al.}(2021)\citenamefont {Lyu},
  \citenamefont {Jiang}, \citenamefont {Wang}, \citenamefont {Xiao},
  \citenamefont {Dong}, \citenamefont {Chen}, \citenamefont {Milošević},
  \citenamefont {Wang}, \citenamefont {Divan}, \citenamefont {Pearson},
  \citenamefont {Wu}, \citenamefont {Peeters},\ and\ \citenamefont
  {Kwok}}]{Lyu_NC_2021}%
  \BibitemOpen
  \bibfield  {author} {\bibinfo {author} {\bibfnamefont {Y.-Y.}\ \bibnamefont
  {Lyu}}, \bibinfo {author} {\bibfnamefont {J.}~\bibnamefont {Jiang}}, \bibinfo
  {author} {\bibfnamefont {Y.-L.}\ \bibnamefont {Wang}}, \bibinfo {author}
  {\bibfnamefont {Z.-L.}\ \bibnamefont {Xiao}}, \bibinfo {author}
  {\bibfnamefont {S.}~\bibnamefont {Dong}}, \bibinfo {author} {\bibfnamefont
  {Q.-H.}\ \bibnamefont {Chen}}, \bibinfo {author} {\bibfnamefont {M.~V.}\
  \bibnamefont {Milošević}}, \bibinfo {author} {\bibfnamefont
  {H.}~\bibnamefont {Wang}}, \bibinfo {author} {\bibfnamefont {R.}~\bibnamefont
  {Divan}}, \bibinfo {author} {\bibfnamefont {J.~E.}\ \bibnamefont {Pearson}},
  \bibinfo {author} {\bibfnamefont {P.}~\bibnamefont {Wu}}, \bibinfo {author}
  {\bibfnamefont {F.~M.}\ \bibnamefont {Peeters}},\ and\ \bibinfo {author}
  {\bibfnamefont {W.-K.}\ \bibnamefont {Kwok}},\ }\bibfield  {title} {\bibinfo
  {title} {Superconducting diode effect via conformal-mapped nanoholes},\
  }\href {https://doi.org/10.1038/s41467-021-23077-0} {\bibfield  {journal}
  {\bibinfo  {journal} {Nat. Commun.}\ }\textbf {\bibinfo {volume} {12}},\
  \bibinfo {pages} {2703} (\bibinfo {year} {2021})}\BibitemShut {NoStop}%
\bibitem [{\citenamefont {Turini}\ \emph {et~al.}(2022)\citenamefont {Turini},
  \citenamefont {Salimian}, \citenamefont {Carrega}, \citenamefont {Iorio},
  \citenamefont {Strambini}, \citenamefont {Giazotto}, \citenamefont {Zannier},
  \citenamefont {Sorba},\ and\ \citenamefont {Heun}}]{Turini_NL_2022}%
  \BibitemOpen
  \bibfield  {author} {\bibinfo {author} {\bibfnamefont {B.}~\bibnamefont
  {Turini}}, \bibinfo {author} {\bibfnamefont {S.}~\bibnamefont {Salimian}},
  \bibinfo {author} {\bibfnamefont {M.}~\bibnamefont {Carrega}}, \bibinfo
  {author} {\bibfnamefont {A.}~\bibnamefont {Iorio}}, \bibinfo {author}
  {\bibfnamefont {E.}~\bibnamefont {Strambini}}, \bibinfo {author}
  {\bibfnamefont {F.}~\bibnamefont {Giazotto}}, \bibinfo {author}
  {\bibfnamefont {V.}~\bibnamefont {Zannier}}, \bibinfo {author} {\bibfnamefont
  {L.}~\bibnamefont {Sorba}},\ and\ \bibinfo {author} {\bibfnamefont
  {S.}~\bibnamefont {Heun}},\ }\bibfield  {title} {\bibinfo {title} {Josephson
  diode effect in high-mobility insb nanoflags},\ }\href
  {https://doi.org/10.1021/acs.nanolett.2c02899} {\bibfield  {journal}
  {\bibinfo  {journal} {Nano Lett.}\ }\textbf {\bibinfo {volume} {22}},\
  \bibinfo {pages} {8502} (\bibinfo {year} {2022})}\BibitemShut {NoStop}%
\bibitem [{\citenamefont {Strambini}\ \emph {et~al.}(2022)\citenamefont
  {Strambini}, \citenamefont {Spies}, \citenamefont {Ligato}, \citenamefont
  {Ilic}, \citenamefont {Rouco}, \citenamefont {Gonzalez-Orellana},
  \citenamefont {Ilyn}, \citenamefont {Rogero}, \citenamefont {Bergeret},
  \citenamefont {Moodera}, \citenamefont {Virtanen}, \citenamefont {Heikkila},\
  and\ \citenamefont {Giazotto}}]{Strambini_NC_2022}%
  \BibitemOpen
  \bibfield  {author} {\bibinfo {author} {\bibfnamefont {E.}~\bibnamefont
  {Strambini}}, \bibinfo {author} {\bibfnamefont {M.}~\bibnamefont {Spies}},
  \bibinfo {author} {\bibfnamefont {N.}~\bibnamefont {Ligato}}, \bibinfo
  {author} {\bibfnamefont {S.}~\bibnamefont {Ilic}}, \bibinfo {author}
  {\bibfnamefont {M.}~\bibnamefont {Rouco}}, \bibinfo {author} {\bibfnamefont
  {C.}~\bibnamefont {Gonzalez-Orellana}}, \bibinfo {author} {\bibfnamefont
  {M.}~\bibnamefont {Ilyn}}, \bibinfo {author} {\bibfnamefont {C.}~\bibnamefont
  {Rogero}}, \bibinfo {author} {\bibfnamefont {F.~S.}\ \bibnamefont
  {Bergeret}}, \bibinfo {author} {\bibfnamefont {J.~S.}\ \bibnamefont
  {Moodera}}, \bibinfo {author} {\bibfnamefont {P.}~\bibnamefont {Virtanen}},
  \bibinfo {author} {\bibfnamefont {T.~T.}\ \bibnamefont {Heikkila}},\ and\
  \bibinfo {author} {\bibfnamefont {F.}~\bibnamefont {Giazotto}},\ }\bibfield
  {title} {\bibinfo {title} {Superconducting spintronic tunnel diode},\ }\href
  {https://doi.org/10.1038/s41467-022-29990-2} {\bibfield  {journal} {\bibinfo
  {journal} {Nat. Commun.}\ }\textbf {\bibinfo {volume} {13}},\ \bibinfo
  {pages} {2431} (\bibinfo {year} {2022})}\BibitemShut {NoStop}%
\bibitem [{\citenamefont {Golod}\ and\ \citenamefont
  {Krasnov}(2022)}]{Golod_NC_2022}%
  \BibitemOpen
  \bibfield  {author} {\bibinfo {author} {\bibfnamefont {T.}~\bibnamefont
  {Golod}}\ and\ \bibinfo {author} {\bibfnamefont {V.~M.}\ \bibnamefont
  {Krasnov}},\ }\bibfield  {title} {\bibinfo {title} {Demonstration of a
  superconducting diode-with-memory, operational at zero magnetic field with
  switchable nonreciprocity},\ }\href
  {https://doi.org/10.1038/s41467-022-31256-w} {\bibfield  {journal} {\bibinfo
  {journal} {Nat. Commun.}\ }\textbf {\bibinfo {volume} {13}},\ \bibinfo
  {pages} {3658} (\bibinfo {year} {2022})}\BibitemShut {NoStop}%
\bibitem [{\citenamefont {Jeon}\ \emph {et~al.}(2022)\citenamefont {Jeon},
  \citenamefont {Kim}, \citenamefont {Yoon}, \citenamefont {Jeon},
  \citenamefont {Han}, \citenamefont {Cottet}, \citenamefont {Kontos},\ and\
  \citenamefont {Parkin}}]{Jeon_NM_2022}%
  \BibitemOpen
  \bibfield  {author} {\bibinfo {author} {\bibfnamefont {K.-R.}\ \bibnamefont
  {Jeon}}, \bibinfo {author} {\bibfnamefont {J.-K.}\ \bibnamefont {Kim}},
  \bibinfo {author} {\bibfnamefont {J.}~\bibnamefont {Yoon}}, \bibinfo {author}
  {\bibfnamefont {J.-C.}\ \bibnamefont {Jeon}}, \bibinfo {author}
  {\bibfnamefont {H.}~\bibnamefont {Han}}, \bibinfo {author} {\bibfnamefont
  {A.}~\bibnamefont {Cottet}}, \bibinfo {author} {\bibfnamefont
  {T.}~\bibnamefont {Kontos}},\ and\ \bibinfo {author} {\bibfnamefont
  {S.~S.~P.}\ \bibnamefont {Parkin}},\ }\bibfield  {title} {\bibinfo {title}
  {Zero-field polarity-reversible josephson supercurrent diodes enabled by a
  proximity-magnetized pt barrier},\ }\href
  {https://doi.org/10.1038/s41563-022-01300-7} {\bibfield  {journal} {\bibinfo
  {journal} {Nat. Mater.}\ }\textbf {\bibinfo {volume} {21}},\ \bibinfo {pages}
  {1008} (\bibinfo {year} {2022})}\BibitemShut {NoStop}%
\bibitem [{\citenamefont {Narita}\ \emph {et~al.}(2022)\citenamefont {Narita},
  \citenamefont {Ishizuka}, \citenamefont {Kawarazaki}, \citenamefont {Kan},
  \citenamefont {Shiota}, \citenamefont {Moriyama}, \citenamefont {Shimakawa},
  \citenamefont {Ognev}, \citenamefont {Samardak}, \citenamefont {Yanase},\
  and\ \citenamefont {Ono}}]{Narita_NN_2022}%
  \BibitemOpen
  \bibfield  {author} {\bibinfo {author} {\bibfnamefont {H.}~\bibnamefont
  {Narita}}, \bibinfo {author} {\bibfnamefont {J.}~\bibnamefont {Ishizuka}},
  \bibinfo {author} {\bibfnamefont {R.}~\bibnamefont {Kawarazaki}}, \bibinfo
  {author} {\bibfnamefont {D.}~\bibnamefont {Kan}}, \bibinfo {author}
  {\bibfnamefont {Y.}~\bibnamefont {Shiota}}, \bibinfo {author} {\bibfnamefont
  {T.}~\bibnamefont {Moriyama}}, \bibinfo {author} {\bibfnamefont
  {Y.}~\bibnamefont {Shimakawa}}, \bibinfo {author} {\bibfnamefont {A.~V.}\
  \bibnamefont {Ognev}}, \bibinfo {author} {\bibfnamefont {A.~S.}\ \bibnamefont
  {Samardak}}, \bibinfo {author} {\bibfnamefont {Y.}~\bibnamefont {Yanase}},\
  and\ \bibinfo {author} {\bibfnamefont {T.}~\bibnamefont {Ono}},\ }\bibfield
  {title} {\bibinfo {title} {Field-free superconducting diode effect in
  noncentrosymmetric superconductor/ferromagnet multilayers},\ }\href
  {https://doi.org/10.1038/s41565-022-01159-4} {\bibfield  {journal} {\bibinfo
  {journal} {Nat. Nanotechnol.}\ }\textbf {\bibinfo {volume} {17}},\ \bibinfo
  {pages} {823} (\bibinfo {year} {2022})}\BibitemShut {NoStop}%
\bibitem [{\citenamefont {Lin}\ \emph {et~al.}(2022)\citenamefont {Lin},
  \citenamefont {Siriviboon}, \citenamefont {Scammell}, \citenamefont {Liu},
  \citenamefont {Rhodes}, \citenamefont {Watanabe}, \citenamefont {Taniguchi},
  \citenamefont {Hone}, \citenamefont {Scheurer},\ and\ \citenamefont
  {Li}}]{Lin_NP_2022}%
  \BibitemOpen
  \bibfield  {author} {\bibinfo {author} {\bibfnamefont {J.-X.}\ \bibnamefont
  {Lin}}, \bibinfo {author} {\bibfnamefont {P.}~\bibnamefont {Siriviboon}},
  \bibinfo {author} {\bibfnamefont {H.~D.}\ \bibnamefont {Scammell}}, \bibinfo
  {author} {\bibfnamefont {S.}~\bibnamefont {Liu}}, \bibinfo {author}
  {\bibfnamefont {D.}~\bibnamefont {Rhodes}}, \bibinfo {author} {\bibfnamefont
  {K.}~\bibnamefont {Watanabe}}, \bibinfo {author} {\bibfnamefont
  {T.}~\bibnamefont {Taniguchi}}, \bibinfo {author} {\bibfnamefont
  {J.}~\bibnamefont {Hone}}, \bibinfo {author} {\bibfnamefont {M.~S.}\
  \bibnamefont {Scheurer}},\ and\ \bibinfo {author} {\bibfnamefont {J.~I.~A.}\
  \bibnamefont {Li}},\ }\bibfield  {title} {\bibinfo {title} {Zero-field
  superconducting diode effect in small-twist-angle trilayer graphene},\ }\href
  {https://doi.org/10.1038/s41567-022-01700-1} {\bibfield  {journal} {\bibinfo
  {journal} {Nat. Phys.}\ }\textbf {\bibinfo {volume} {18}},\ \bibinfo {pages}
  {1221} (\bibinfo {year} {2022})}\BibitemShut {NoStop}%
\bibitem [{\citenamefont {Pal}\ \emph {et~al.}(2022)\citenamefont {Pal},
  \citenamefont {Chakraborty}, \citenamefont {Sivakumar}, \citenamefont
  {Davydova}, \citenamefont {Gopi}, \citenamefont {Pandeya}, \citenamefont
  {Krieger}, \citenamefont {Zhang}, \citenamefont {Date}, \citenamefont {Ju},
  \citenamefont {Yuan}, \citenamefont {Schröter}, \citenamefont {Fu},\ and\
  \citenamefont {Parkin}}]{Pal_NP_2022}%
  \BibitemOpen
  \bibfield  {author} {\bibinfo {author} {\bibfnamefont {B.}~\bibnamefont
  {Pal}}, \bibinfo {author} {\bibfnamefont {A.}~\bibnamefont {Chakraborty}},
  \bibinfo {author} {\bibfnamefont {P.~K.}\ \bibnamefont {Sivakumar}}, \bibinfo
  {author} {\bibfnamefont {M.}~\bibnamefont {Davydova}}, \bibinfo {author}
  {\bibfnamefont {A.~K.}\ \bibnamefont {Gopi}}, \bibinfo {author}
  {\bibfnamefont {A.~K.}\ \bibnamefont {Pandeya}}, \bibinfo {author}
  {\bibfnamefont {J.~A.}\ \bibnamefont {Krieger}}, \bibinfo {author}
  {\bibfnamefont {Y.}~\bibnamefont {Zhang}}, \bibinfo {author} {\bibfnamefont
  {M.}~\bibnamefont {Date}}, \bibinfo {author} {\bibfnamefont {S.}~\bibnamefont
  {Ju}}, \bibinfo {author} {\bibfnamefont {N.}~\bibnamefont {Yuan}}, \bibinfo
  {author} {\bibfnamefont {N.~B.~M.}\ \bibnamefont {Schröter}}, \bibinfo
  {author} {\bibfnamefont {L.}~\bibnamefont {Fu}},\ and\ \bibinfo {author}
  {\bibfnamefont {S.~S.~P.}\ \bibnamefont {Parkin}},\ }\bibfield  {title}
  {\bibinfo {title} {Josephson diode effect from {Cooper} pair momentum in a
  topological semimetal},\ }\href {https://doi.org/10.1038/s41567-022-01699-5}
  {\bibfield  {journal} {\bibinfo  {journal} {Nat. Phys.}\ }\textbf {\bibinfo
  {volume} {18}},\ \bibinfo {pages} {1228} (\bibinfo {year}
  {2022})}\BibitemShut {NoStop}%
\bibitem [{\citenamefont {Wu}\ \emph {et~al.}(2022)\citenamefont {Wu},
  \citenamefont {Wang}, \citenamefont {Xu}, \citenamefont {Sivakumar},
  \citenamefont {Pasco}, \citenamefont {Filippozzi}, \citenamefont {Parkin},
  \citenamefont {Zeng}, \citenamefont {McQueen},\ and\ \citenamefont
  {Ali}}]{Wu_N_2022}%
  \BibitemOpen
  \bibfield  {author} {\bibinfo {author} {\bibfnamefont {H.}~\bibnamefont
  {Wu}}, \bibinfo {author} {\bibfnamefont {Y.}~\bibnamefont {Wang}}, \bibinfo
  {author} {\bibfnamefont {Y.}~\bibnamefont {Xu}}, \bibinfo {author}
  {\bibfnamefont {P.~K.}\ \bibnamefont {Sivakumar}}, \bibinfo {author}
  {\bibfnamefont {C.}~\bibnamefont {Pasco}}, \bibinfo {author} {\bibfnamefont
  {U.}~\bibnamefont {Filippozzi}}, \bibinfo {author} {\bibfnamefont {S.~S.~P.}\
  \bibnamefont {Parkin}}, \bibinfo {author} {\bibfnamefont {Y.-J.}\
  \bibnamefont {Zeng}}, \bibinfo {author} {\bibfnamefont {T.}~\bibnamefont
  {McQueen}},\ and\ \bibinfo {author} {\bibfnamefont {M.~N.}\ \bibnamefont
  {Ali}},\ }\bibfield  {title} {\bibinfo {title} {The field-free {{Josephson}}
  diode in a van der {{Waals}} heterostructure},\ }\href
  {https://doi.org/10.1038/s41586-022-04504-8} {\bibfield  {journal} {\bibinfo
  {journal} {Nature}\ }\textbf {\bibinfo {volume} {604}},\ \bibinfo {pages}
  {653} (\bibinfo {year} {2022})}\BibitemShut {NoStop}%
\bibitem [{\citenamefont {Jiang}\ \emph {et~al.}(2022)\citenamefont {Jiang},
  \citenamefont {Milošević}, \citenamefont {Wang}, \citenamefont {Xiao},
  \citenamefont {Peeters},\ and\ \citenamefont {Chen}}]{Jiang_PRA_2022}%
  \BibitemOpen
  \bibfield  {author} {\bibinfo {author} {\bibfnamefont {J.}~\bibnamefont
  {Jiang}}, \bibinfo {author} {\bibfnamefont {M.}~\bibnamefont {Milošević}},
  \bibinfo {author} {\bibfnamefont {Y.-L.}\ \bibnamefont {Wang}}, \bibinfo
  {author} {\bibfnamefont {Z.-L.}\ \bibnamefont {Xiao}}, \bibinfo {author}
  {\bibfnamefont {F.}~\bibnamefont {Peeters}},\ and\ \bibinfo {author}
  {\bibfnamefont {Q.-H.}\ \bibnamefont {Chen}},\ }\bibfield  {title} {\bibinfo
  {title} {Field-{Free} {Superconducting} {Diode} in a {Magnetically}
  {Nanostructured} {Superconductor}},\ }\href
  {https://doi.org/10.1103/PhysRevApplied.18.034064} {\bibfield  {journal}
  {\bibinfo  {journal} {Phys. Rev. Applied}\ }\textbf {\bibinfo {volume}
  {18}},\ \bibinfo {pages} {034064} (\bibinfo {year} {2022})}\BibitemShut
  {NoStop}%
\bibitem [{\citenamefont {Chiles}\ \emph {et~al.}(2023)\citenamefont {Chiles},
  \citenamefont {Arnault}, \citenamefont {Chen}, \citenamefont {Larson},
  \citenamefont {Zhao}, \citenamefont {Watanabe}, \citenamefont {Taniguchi},
  \citenamefont {Amet},\ and\ \citenamefont {Finkelstein}}]{Chiles_NL_2023}%
  \BibitemOpen
  \bibfield  {author} {\bibinfo {author} {\bibfnamefont {J.}~\bibnamefont
  {Chiles}}, \bibinfo {author} {\bibfnamefont {E.~G.}\ \bibnamefont {Arnault}},
  \bibinfo {author} {\bibfnamefont {C.-C.}\ \bibnamefont {Chen}}, \bibinfo
  {author} {\bibfnamefont {T.~F.~Q.}\ \bibnamefont {Larson}}, \bibinfo {author}
  {\bibfnamefont {L.}~\bibnamefont {Zhao}}, \bibinfo {author} {\bibfnamefont
  {K.}~\bibnamefont {Watanabe}}, \bibinfo {author} {\bibfnamefont
  {T.}~\bibnamefont {Taniguchi}}, \bibinfo {author} {\bibfnamefont
  {F.}~\bibnamefont {Amet}},\ and\ \bibinfo {author} {\bibfnamefont
  {G.}~\bibnamefont {Finkelstein}},\ }\bibfield  {title} {\bibinfo {title}
  {Nonreciprocal supercurrents in a field-free graphene josephson triode},\
  }\href {https://doi.org/10.1021/acs.nanolett.3c01276} {\bibfield  {journal}
  {\bibinfo  {journal} {Nano Lett.}\ }\textbf {\bibinfo {volume} {23}},\
  \bibinfo {pages} {5257} (\bibinfo {year} {2023})}\BibitemShut {NoStop}%
\bibitem [{\citenamefont {Gupta}\ \emph {et~al.}(2023)\citenamefont {Gupta},
  \citenamefont {Graziano}, \citenamefont {Pendharkar}, \citenamefont {Dong},
  \citenamefont {Dempsey}, \citenamefont {Palmstrom},\ and\ \citenamefont
  {Pribiag}}]{Gupta_NC_2023}%
  \BibitemOpen
  \bibfield  {author} {\bibinfo {author} {\bibfnamefont {M.}~\bibnamefont
  {Gupta}}, \bibinfo {author} {\bibfnamefont {G.~V.}\ \bibnamefont {Graziano}},
  \bibinfo {author} {\bibfnamefont {M.}~\bibnamefont {Pendharkar}}, \bibinfo
  {author} {\bibfnamefont {J.~T.}\ \bibnamefont {Dong}}, \bibinfo {author}
  {\bibfnamefont {C.~P.}\ \bibnamefont {Dempsey}}, \bibinfo {author}
  {\bibfnamefont {C.}~\bibnamefont {Palmstrom}},\ and\ \bibinfo {author}
  {\bibfnamefont {V.~S.}\ \bibnamefont {Pribiag}},\ }\bibfield  {title}
  {\bibinfo {title} {Gate-tunable superconducting diode effect in a
  three-terminal josephson device},\ }\href
  {https://doi.org/10.1038/s41467-023-38856-0} {\bibfield  {journal} {\bibinfo
  {journal} {Nat. Commun.}\ }\textbf {\bibinfo {volume} {14}},\ \bibinfo
  {pages} {3078} (\bibinfo {year} {2023})}\BibitemShut {NoStop}%
\bibitem [{\citenamefont {Diez-Merida}\ \emph {et~al.}(2023)\citenamefont
  {Diez-Merida}, \citenamefont {Diez-Carlon}, \citenamefont {Yang},
  \citenamefont {Xie}, \citenamefont {Gao}, \citenamefont {Senior},
  \citenamefont {Watanabe}, \citenamefont {Taniguchi}, \citenamefont {Lu},
  \citenamefont {Higginbotham}, \citenamefont {Law},\ and\ \citenamefont
  {Efetov}}]{DiezMerida_NC_2023}%
  \BibitemOpen
  \bibfield  {author} {\bibinfo {author} {\bibfnamefont {J.}~\bibnamefont
  {Diez-Merida}}, \bibinfo {author} {\bibfnamefont {A.}~\bibnamefont
  {Diez-Carlon}}, \bibinfo {author} {\bibfnamefont {S.~Y.}\ \bibnamefont
  {Yang}}, \bibinfo {author} {\bibfnamefont {Y.~M.}\ \bibnamefont {Xie}},
  \bibinfo {author} {\bibfnamefont {X.~J.}\ \bibnamefont {Gao}}, \bibinfo
  {author} {\bibfnamefont {J.}~\bibnamefont {Senior}}, \bibinfo {author}
  {\bibfnamefont {K.}~\bibnamefont {Watanabe}}, \bibinfo {author}
  {\bibfnamefont {T.}~\bibnamefont {Taniguchi}}, \bibinfo {author}
  {\bibfnamefont {X.}~\bibnamefont {Lu}}, \bibinfo {author} {\bibfnamefont
  {A.~P.}\ \bibnamefont {Higginbotham}}, \bibinfo {author} {\bibfnamefont
  {K.~T.}\ \bibnamefont {Law}},\ and\ \bibinfo {author} {\bibfnamefont {D.~K.}\
  \bibnamefont {Efetov}},\ }\bibfield  {title} {\bibinfo {title}
  {Symmetry-broken josephson junctions and superconducting diodes in
  magic-angle twisted bilayer graphene},\ }\href
  {https://doi.org/10.1038/s41467-023-38005-7} {\bibfield  {journal} {\bibinfo
  {journal} {Nat. Commun.}\ }\textbf {\bibinfo {volume} {14}},\ \bibinfo
  {pages} {2396} (\bibinfo {year} {2023})}\BibitemShut {NoStop}%
\bibitem [{\citenamefont {Gutfreund}\ \emph {et~al.}(2023)\citenamefont
  {Gutfreund}, \citenamefont {Matsuki}, \citenamefont {Plastovets},
  \citenamefont {Noah}, \citenamefont {Gorzawski}, \citenamefont {Fridman},
  \citenamefont {Yang}, \citenamefont {Buzdin}, \citenamefont {Millo},
  \citenamefont {Robinson},\ and\ \citenamefont {Anahory}}]{Gutfreund_NC_2023}%
  \BibitemOpen
  \bibfield  {author} {\bibinfo {author} {\bibfnamefont {A.}~\bibnamefont
  {Gutfreund}}, \bibinfo {author} {\bibfnamefont {H.}~\bibnamefont {Matsuki}},
  \bibinfo {author} {\bibfnamefont {V.}~\bibnamefont {Plastovets}}, \bibinfo
  {author} {\bibfnamefont {A.}~\bibnamefont {Noah}}, \bibinfo {author}
  {\bibfnamefont {L.}~\bibnamefont {Gorzawski}}, \bibinfo {author}
  {\bibfnamefont {N.}~\bibnamefont {Fridman}}, \bibinfo {author} {\bibfnamefont
  {G.}~\bibnamefont {Yang}}, \bibinfo {author} {\bibfnamefont {A.}~\bibnamefont
  {Buzdin}}, \bibinfo {author} {\bibfnamefont {O.}~\bibnamefont {Millo}},
  \bibinfo {author} {\bibfnamefont {J.~W.~A.}\ \bibnamefont {Robinson}},\ and\
  \bibinfo {author} {\bibfnamefont {Y.}~\bibnamefont {Anahory}},\ }\bibfield
  {title} {\bibinfo {title} {Direct observation of a superconducting vortex
  diode},\ }\href {https://doi.org/10.1038/s41467-023-37294-2} {\bibfield
  {journal} {\bibinfo  {journal} {Nat. Commun.}\ }\textbf {\bibinfo {volume}
  {14}},\ \bibinfo {pages} {1630} (\bibinfo {year} {2023})}\BibitemShut
  {NoStop}%
\bibitem [{\citenamefont {Matsuo}\ \emph {et~al.}(2023)\citenamefont {Matsuo},
  \citenamefont {Imoto}, \citenamefont {Yokoyama}, \citenamefont {Sato},
  \citenamefont {Lindemann}, \citenamefont {Gronin}, \citenamefont {Gardner},
  \citenamefont {Manfra},\ and\ \citenamefont {Tarucha}}]{Matsuo_NP_2023}%
  \BibitemOpen
  \bibfield  {author} {\bibinfo {author} {\bibfnamefont {S.}~\bibnamefont
  {Matsuo}}, \bibinfo {author} {\bibfnamefont {T.}~\bibnamefont {Imoto}},
  \bibinfo {author} {\bibfnamefont {T.}~\bibnamefont {Yokoyama}}, \bibinfo
  {author} {\bibfnamefont {Y.}~\bibnamefont {Sato}}, \bibinfo {author}
  {\bibfnamefont {T.}~\bibnamefont {Lindemann}}, \bibinfo {author}
  {\bibfnamefont {S.}~\bibnamefont {Gronin}}, \bibinfo {author} {\bibfnamefont
  {G.~C.}\ \bibnamefont {Gardner}}, \bibinfo {author} {\bibfnamefont {M.~J.}\
  \bibnamefont {Manfra}},\ and\ \bibinfo {author} {\bibfnamefont
  {S.}~\bibnamefont {Tarucha}},\ }\bibfield  {title} {\bibinfo {title}
  {Josephson diode effect derived from short-range coherent coupling},\ }\href
  {https://doi.org/10.1038/s41567-023-02144-x} {\bibfield  {journal} {\bibinfo
  {journal} {Nat. Phys.}\ }\textbf {\bibinfo {volume} {19}},\ \bibinfo {pages}
  {1636} (\bibinfo {year} {2023})}\BibitemShut {NoStop}%
\bibitem [{\citenamefont {Trahms}\ \emph {et~al.}(2023)\citenamefont {Trahms},
  \citenamefont {Melischek}, \citenamefont {Steiner}, \citenamefont {Mahendru},
  \citenamefont {Tamir}, \citenamefont {Bogdanoff}, \citenamefont {Peters},
  \citenamefont {Reecht}, \citenamefont {Winkelmann}, \citenamefont {von
  Oppen},\ and\ \citenamefont {Franke}}]{Trahms_N_2023}%
  \BibitemOpen
  \bibfield  {author} {\bibinfo {author} {\bibfnamefont {M.}~\bibnamefont
  {Trahms}}, \bibinfo {author} {\bibfnamefont {L.}~\bibnamefont {Melischek}},
  \bibinfo {author} {\bibfnamefont {J.~F.}\ \bibnamefont {Steiner}}, \bibinfo
  {author} {\bibfnamefont {B.}~\bibnamefont {Mahendru}}, \bibinfo {author}
  {\bibfnamefont {I.}~\bibnamefont {Tamir}}, \bibinfo {author} {\bibfnamefont
  {N.}~\bibnamefont {Bogdanoff}}, \bibinfo {author} {\bibfnamefont
  {O.}~\bibnamefont {Peters}}, \bibinfo {author} {\bibfnamefont
  {G.}~\bibnamefont {Reecht}}, \bibinfo {author} {\bibfnamefont {C.~B.}\
  \bibnamefont {Winkelmann}}, \bibinfo {author} {\bibfnamefont
  {F.}~\bibnamefont {von Oppen}},\ and\ \bibinfo {author} {\bibfnamefont
  {K.~J.}\ \bibnamefont {Franke}},\ }\bibfield  {title} {\bibinfo {title}
  {Diode effect in {Josephson} junctions with a single magnetic atom},\ }\href
  {https://doi.org/10.1038/s41586-023-05743-z} {\bibfield  {journal} {\bibinfo
  {journal} {Nature}\ }\textbf {\bibinfo {volume} {615}},\ \bibinfo {pages}
  {628} (\bibinfo {year} {2023})}\BibitemShut {NoStop}%
\bibitem [{\citenamefont {Hou}\ \emph {et~al.}(2023)\citenamefont {Hou},
  \citenamefont {Nichele}, \citenamefont {Chi}, \citenamefont {Lodesani},
  \citenamefont {Wu}, \citenamefont {Ritter}, \citenamefont {Haxell},
  \citenamefont {Davydova}, \citenamefont {Ilić}, \citenamefont
  {Glezakou-Elbert}, \citenamefont {Varambally}, \citenamefont {Bergeret},
  \citenamefont {Kamra}, \citenamefont {Fu}, \citenamefont {Lee},\ and\
  \citenamefont {Moodera}}]{Hou_PRL_2023}%
  \BibitemOpen
  \bibfield  {author} {\bibinfo {author} {\bibfnamefont {Y.}~\bibnamefont
  {Hou}}, \bibinfo {author} {\bibfnamefont {F.}~\bibnamefont {Nichele}},
  \bibinfo {author} {\bibfnamefont {H.}~\bibnamefont {Chi}}, \bibinfo {author}
  {\bibfnamefont {A.}~\bibnamefont {Lodesani}}, \bibinfo {author}
  {\bibfnamefont {Y.}~\bibnamefont {Wu}}, \bibinfo {author} {\bibfnamefont
  {M.~F.}\ \bibnamefont {Ritter}}, \bibinfo {author} {\bibfnamefont
  {D.}~\bibnamefont {Haxell}}, \bibinfo {author} {\bibfnamefont
  {M.}~\bibnamefont {Davydova}}, \bibinfo {author} {\bibfnamefont
  {S.}~\bibnamefont {Ilić}}, \bibinfo {author} {\bibfnamefont
  {O.}~\bibnamefont {Glezakou-Elbert}}, \bibinfo {author} {\bibfnamefont
  {A.}~\bibnamefont {Varambally}}, \bibinfo {author} {\bibfnamefont {F.~S.}\
  \bibnamefont {Bergeret}}, \bibinfo {author} {\bibfnamefont {A.}~\bibnamefont
  {Kamra}}, \bibinfo {author} {\bibfnamefont {L.}~\bibnamefont {Fu}}, \bibinfo
  {author} {\bibfnamefont {P.~A.}\ \bibnamefont {Lee}},\ and\ \bibinfo {author}
  {\bibfnamefont {J.~S.}\ \bibnamefont {Moodera}},\ }\bibfield  {title}
  {\bibinfo {title} {Ubiquitous {Superconducting} {Diode} {Effect} in
  {Superconductor} {Thin} {Films}},\ }\href
  {https://doi.org/10.1103/PhysRevLett.131.027001} {\bibfield  {journal}
  {\bibinfo  {journal} {Phys. Rev. Lett.}\ }\textbf {\bibinfo {volume} {131}},\
  \bibinfo {pages} {027001} (\bibinfo {year} {2023})}\BibitemShut {NoStop}%
\bibitem [{\citenamefont {Zhao}\ \emph {et~al.}(2023)\citenamefont {Zhao},
  \citenamefont {Cui}, \citenamefont {Volkov}, \citenamefont {Yoo},
  \citenamefont {Lee}, \citenamefont {Gardener}, \citenamefont {Akey},
  \citenamefont {Engelke}, \citenamefont {Ronen}, \citenamefont {Zhong},
  \citenamefont {Gu}, \citenamefont {Plugge}, \citenamefont {Tummuru},
  \citenamefont {Kim}, \citenamefont {Franz}, \citenamefont {Pixley},
  \citenamefont {Poccia},\ and\ \citenamefont {Kim}}]{Zhao_S_2023}%
  \BibitemOpen
  \bibfield  {author} {\bibinfo {author} {\bibfnamefont {S.~Y.~F.}\
  \bibnamefont {Zhao}}, \bibinfo {author} {\bibfnamefont {X.}~\bibnamefont
  {Cui}}, \bibinfo {author} {\bibfnamefont {P.~A.}\ \bibnamefont {Volkov}},
  \bibinfo {author} {\bibfnamefont {H.}~\bibnamefont {Yoo}}, \bibinfo {author}
  {\bibfnamefont {S.}~\bibnamefont {Lee}}, \bibinfo {author} {\bibfnamefont
  {J.~A.}\ \bibnamefont {Gardener}}, \bibinfo {author} {\bibfnamefont {A.~J.}\
  \bibnamefont {Akey}}, \bibinfo {author} {\bibfnamefont {R.}~\bibnamefont
  {Engelke}}, \bibinfo {author} {\bibfnamefont {Y.}~\bibnamefont {Ronen}},
  \bibinfo {author} {\bibfnamefont {R.}~\bibnamefont {Zhong}}, \bibinfo
  {author} {\bibfnamefont {G.}~\bibnamefont {Gu}}, \bibinfo {author}
  {\bibfnamefont {S.}~\bibnamefont {Plugge}}, \bibinfo {author} {\bibfnamefont
  {T.}~\bibnamefont {Tummuru}}, \bibinfo {author} {\bibfnamefont
  {M.}~\bibnamefont {Kim}}, \bibinfo {author} {\bibfnamefont {M.}~\bibnamefont
  {Franz}}, \bibinfo {author} {\bibfnamefont {J.~H.}\ \bibnamefont {Pixley}},
  \bibinfo {author} {\bibfnamefont {N.}~\bibnamefont {Poccia}},\ and\ \bibinfo
  {author} {\bibfnamefont {P.}~\bibnamefont {Kim}},\ }\bibfield  {title}
  {\bibinfo {title} {Time-reversal symmetry breaking superconductivity between
  twisted cuprate superconductors},\ }\href
  {https://doi.org/10.1126/science.abl8371} {\bibfield  {journal} {\bibinfo
  {journal} {Science}\ }\textbf {\bibinfo {volume} {382}},\ \bibinfo {pages}
  {1422} (\bibinfo {year} {2023})}\BibitemShut {NoStop}%
\bibitem [{\citenamefont {Li}\ \emph {et~al.}(2024{\natexlab{a}})\citenamefont
  {Li}, \citenamefont {Deng}, \citenamefont {Hu}, \citenamefont {Zhu},
  \citenamefont {Yang}, \citenamefont {Tian}, \citenamefont {Wang},
  \citenamefont {Yue}, \citenamefont {Wu}, \citenamefont {Liu},\ and\
  \citenamefont {Renshaw~Wang}}]{Li_AN_2024}%
  \BibitemOpen
  \bibfield  {author} {\bibinfo {author} {\bibfnamefont {S.}~\bibnamefont
  {Li}}, \bibinfo {author} {\bibfnamefont {Y.}~\bibnamefont {Deng}}, \bibinfo
  {author} {\bibfnamefont {D.}~\bibnamefont {Hu}}, \bibinfo {author}
  {\bibfnamefont {C.}~\bibnamefont {Zhu}}, \bibinfo {author} {\bibfnamefont
  {Z.}~\bibnamefont {Yang}}, \bibinfo {author} {\bibfnamefont {W.}~\bibnamefont
  {Tian}}, \bibinfo {author} {\bibfnamefont {X.}~\bibnamefont {Wang}}, \bibinfo
  {author} {\bibfnamefont {M.}~\bibnamefont {Yue}}, \bibinfo {author}
  {\bibfnamefont {Q.}~\bibnamefont {Wu}}, \bibinfo {author} {\bibfnamefont
  {Z.}~\bibnamefont {Liu}},\ and\ \bibinfo {author} {\bibfnamefont
  {X.}~\bibnamefont {Renshaw~Wang}},\ }\bibfield  {title} {\bibinfo {title}
  {Field-{Free} {Superconducting} {Diode} {Effect} and {Magnetochiral}
  {Anisotropy} in {FeTe0}.{7Se0}.3 {Junctions} with the {Inherent} {Asymmetric}
  {Barrier}},\ }\href {https://doi.org/10.1021/acsnano.4c07951} {\bibfield
  {journal} {\bibinfo  {journal} {ACS Nano}\ }\textbf {\bibinfo {volume}
  {18}},\ \bibinfo {pages} {31076} (\bibinfo {year}
  {2024}{\natexlab{a}})}\BibitemShut {NoStop}%
\bibitem [{\citenamefont {Li}\ \emph {et~al.}(2024{\natexlab{b}})\citenamefont
  {Li}, \citenamefont {Yan}, \citenamefont {Hong}, \citenamefont {Sheng},
  \citenamefont {Wang}, \citenamefont {Dou}, \citenamefont {Guo}, \citenamefont
  {Shi}, \citenamefont {Su}, \citenamefont {Lyu}, \citenamefont {Qian},
  \citenamefont {Liu}, \citenamefont {Qu}, \citenamefont {Jiang}, \citenamefont
  {Wang}, \citenamefont {Shi}, \citenamefont {Xu}, \citenamefont {Hu},
  \citenamefont {Lu},\ and\ \citenamefont {Shen}}]{Li_NC_2024}%
  \BibitemOpen
  \bibfield  {author} {\bibinfo {author} {\bibfnamefont {Y.}~\bibnamefont
  {Li}}, \bibinfo {author} {\bibfnamefont {D.}~\bibnamefont {Yan}}, \bibinfo
  {author} {\bibfnamefont {Y.}~\bibnamefont {Hong}}, \bibinfo {author}
  {\bibfnamefont {H.}~\bibnamefont {Sheng}}, \bibinfo {author} {\bibfnamefont
  {A.}~\bibnamefont {Wang}}, \bibinfo {author} {\bibfnamefont {Z.}~\bibnamefont
  {Dou}}, \bibinfo {author} {\bibfnamefont {X.}~\bibnamefont {Guo}}, \bibinfo
  {author} {\bibfnamefont {X.}~\bibnamefont {Shi}}, \bibinfo {author}
  {\bibfnamefont {Z.}~\bibnamefont {Su}}, \bibinfo {author} {\bibfnamefont
  {Z.}~\bibnamefont {Lyu}}, \bibinfo {author} {\bibfnamefont {T.}~\bibnamefont
  {Qian}}, \bibinfo {author} {\bibfnamefont {G.}~\bibnamefont {Liu}}, \bibinfo
  {author} {\bibfnamefont {F.}~\bibnamefont {Qu}}, \bibinfo {author}
  {\bibfnamefont {K.}~\bibnamefont {Jiang}}, \bibinfo {author} {\bibfnamefont
  {Z.}~\bibnamefont {Wang}}, \bibinfo {author} {\bibfnamefont {Y.}~\bibnamefont
  {Shi}}, \bibinfo {author} {\bibfnamefont {Z.-A.}\ \bibnamefont {Xu}},
  \bibinfo {author} {\bibfnamefont {J.}~\bibnamefont {Hu}}, \bibinfo {author}
  {\bibfnamefont {L.}~\bibnamefont {Lu}},\ and\ \bibinfo {author}
  {\bibfnamefont {J.}~\bibnamefont {Shen}},\ }\bibfield  {title} {\bibinfo
  {title} {Interfering josephson diode effect in ta2pd3te5 asymmetric edge
  interferometer},\ }\href {https://doi.org/10.1038/s41467-024-53383-2}
  {\bibfield  {journal} {\bibinfo  {journal} {Nat. Commun.}\ }\textbf {\bibinfo
  {volume} {15}},\ \bibinfo {pages} {9031} (\bibinfo {year}
  {2024}{\natexlab{b}})}\BibitemShut {NoStop}%
\bibitem [{\citenamefont {Reinhardt}\ \emph {et~al.}(2024)\citenamefont
  {Reinhardt}, \citenamefont {Ascherl}, \citenamefont {Costa}, \citenamefont
  {Berger}, \citenamefont {Gronin}, \citenamefont {Gardner}, \citenamefont
  {Lindemann}, \citenamefont {Manfra}, \citenamefont {Fabian}, \citenamefont
  {Kochan}, \citenamefont {Strunk},\ and\ \citenamefont
  {Paradiso}}]{Reinhardt_NC_2024}%
  \BibitemOpen
  \bibfield  {author} {\bibinfo {author} {\bibfnamefont {S.}~\bibnamefont
  {Reinhardt}}, \bibinfo {author} {\bibfnamefont {T.}~\bibnamefont {Ascherl}},
  \bibinfo {author} {\bibfnamefont {A.}~\bibnamefont {Costa}}, \bibinfo
  {author} {\bibfnamefont {J.}~\bibnamefont {Berger}}, \bibinfo {author}
  {\bibfnamefont {S.}~\bibnamefont {Gronin}}, \bibinfo {author} {\bibfnamefont
  {G.~C.}\ \bibnamefont {Gardner}}, \bibinfo {author} {\bibfnamefont
  {T.}~\bibnamefont {Lindemann}}, \bibinfo {author} {\bibfnamefont {M.~J.}\
  \bibnamefont {Manfra}}, \bibinfo {author} {\bibfnamefont {J.}~\bibnamefont
  {Fabian}}, \bibinfo {author} {\bibfnamefont {D.}~\bibnamefont {Kochan}},
  \bibinfo {author} {\bibfnamefont {C.}~\bibnamefont {Strunk}},\ and\ \bibinfo
  {author} {\bibfnamefont {N.}~\bibnamefont {Paradiso}},\ }\bibfield  {title}
  {\bibinfo {title} {Link between supercurrent diode and anomalous josephson
  effect revealed by gate-controlled interferometry},\ }\href
  {https://doi.org/10.1038/s41467-024-48741-z} {\bibfield  {journal} {\bibinfo
  {journal} {Nat. Commun.}\ }\textbf {\bibinfo {volume} {15}},\ \bibinfo
  {pages} {4413} (\bibinfo {year} {2024})}\BibitemShut {NoStop}%
\bibitem [{\citenamefont {Valentini}\ \emph {et~al.}(2024)\citenamefont
  {Valentini}, \citenamefont {Sagi}, \citenamefont {Baghumyan}, \citenamefont
  {de~Gijsel}, \citenamefont {Jung}, \citenamefont {Calcaterra}, \citenamefont
  {Ballabio}, \citenamefont {Aguilera~Servin}, \citenamefont {Aggarwal},
  \citenamefont {Janik}, \citenamefont {Adletzberger}, \citenamefont
  {Seoane~Souto}, \citenamefont {Leijnse}, \citenamefont {Danon}, \citenamefont
  {Schrade}, \citenamefont {Bakkers}, \citenamefont {Chrastina}, \citenamefont
  {Isella},\ and\ \citenamefont {Katsaros}}]{Valentini_NC_2024}%
  \BibitemOpen
  \bibfield  {author} {\bibinfo {author} {\bibfnamefont {M.}~\bibnamefont
  {Valentini}}, \bibinfo {author} {\bibfnamefont {O.}~\bibnamefont {Sagi}},
  \bibinfo {author} {\bibfnamefont {L.}~\bibnamefont {Baghumyan}}, \bibinfo
  {author} {\bibfnamefont {T.}~\bibnamefont {de~Gijsel}}, \bibinfo {author}
  {\bibfnamefont {J.}~\bibnamefont {Jung}}, \bibinfo {author} {\bibfnamefont
  {S.}~\bibnamefont {Calcaterra}}, \bibinfo {author} {\bibfnamefont
  {A.}~\bibnamefont {Ballabio}}, \bibinfo {author} {\bibfnamefont
  {J.}~\bibnamefont {Aguilera~Servin}}, \bibinfo {author} {\bibfnamefont
  {K.}~\bibnamefont {Aggarwal}}, \bibinfo {author} {\bibfnamefont
  {M.}~\bibnamefont {Janik}}, \bibinfo {author} {\bibfnamefont
  {T.}~\bibnamefont {Adletzberger}}, \bibinfo {author} {\bibfnamefont
  {R.}~\bibnamefont {Seoane~Souto}}, \bibinfo {author} {\bibfnamefont
  {M.}~\bibnamefont {Leijnse}}, \bibinfo {author} {\bibfnamefont
  {J.}~\bibnamefont {Danon}}, \bibinfo {author} {\bibfnamefont
  {C.}~\bibnamefont {Schrade}}, \bibinfo {author} {\bibfnamefont
  {E.}~\bibnamefont {Bakkers}}, \bibinfo {author} {\bibfnamefont
  {D.}~\bibnamefont {Chrastina}}, \bibinfo {author} {\bibfnamefont
  {G.}~\bibnamefont {Isella}},\ and\ \bibinfo {author} {\bibfnamefont
  {G.}~\bibnamefont {Katsaros}},\ }\bibfield  {title} {\bibinfo {title}
  {Parity-conserving cooper-pair transport and ideal superconducting diode in
  planar germanium},\ }\href {https://doi.org/10.1038/s41467-023-44114-0}
  {\bibfield  {journal} {\bibinfo  {journal} {Nat. Commun.}\ }\textbf {\bibinfo
  {volume} {15}},\ \bibinfo {pages} {169} (\bibinfo {year} {2024})}\BibitemShut
  {NoStop}%
\bibitem [{\citenamefont {Anh}\ \emph {et~al.}(2024)\citenamefont {Anh},
  \citenamefont {Ishihara}, \citenamefont {Hotta}, \citenamefont {Inagaki},
  \citenamefont {Maki}, \citenamefont {Saeki}, \citenamefont {Kobayashi},\ and\
  \citenamefont {Tanaka}}]{Anh_NC_2024}%
  \BibitemOpen
  \bibfield  {author} {\bibinfo {author} {\bibfnamefont {L.~D.}\ \bibnamefont
  {Anh}}, \bibinfo {author} {\bibfnamefont {K.}~\bibnamefont {Ishihara}},
  \bibinfo {author} {\bibfnamefont {T.}~\bibnamefont {Hotta}}, \bibinfo
  {author} {\bibfnamefont {K.}~\bibnamefont {Inagaki}}, \bibinfo {author}
  {\bibfnamefont {H.}~\bibnamefont {Maki}}, \bibinfo {author} {\bibfnamefont
  {T.}~\bibnamefont {Saeki}}, \bibinfo {author} {\bibfnamefont
  {M.}~\bibnamefont {Kobayashi}},\ and\ \bibinfo {author} {\bibfnamefont
  {M.}~\bibnamefont {Tanaka}},\ }\bibfield  {title} {\bibinfo {title} {Large
  superconducting diode effect in ion-beam patterned sn-based superconductor
  nanowire/topological dirac semimetal planar heterostructures},\ }\href
  {https://doi.org/10.1038/s41467-024-52080-4} {\bibfield  {journal} {\bibinfo
  {journal} {Nat. Commun.}\ }\textbf {\bibinfo {volume} {15}},\ \bibinfo
  {pages} {8014} (\bibinfo {year} {2024})}\BibitemShut {NoStop}%
\bibitem [{\citenamefont {Ghosh}\ \emph {et~al.}(2024)\citenamefont {Ghosh},
  \citenamefont {Patil}, \citenamefont {Basu}, \citenamefont {Kuldeep},
  \citenamefont {Dutta}, \citenamefont {Jangade}, \citenamefont {Kulkarni},
  \citenamefont {Thamizhavel}, \citenamefont {Steiner}, \citenamefont {von
  Oppen},\ and\ \citenamefont {Deshmukh}}]{Ghosh_NM_2024}%
  \BibitemOpen
  \bibfield  {author} {\bibinfo {author} {\bibfnamefont {S.}~\bibnamefont
  {Ghosh}}, \bibinfo {author} {\bibfnamefont {V.}~\bibnamefont {Patil}},
  \bibinfo {author} {\bibfnamefont {A.}~\bibnamefont {Basu}}, \bibinfo {author}
  {\bibnamefont {Kuldeep}}, \bibinfo {author} {\bibfnamefont {A.}~\bibnamefont
  {Dutta}}, \bibinfo {author} {\bibfnamefont {D.~A.}\ \bibnamefont {Jangade}},
  \bibinfo {author} {\bibfnamefont {R.}~\bibnamefont {Kulkarni}}, \bibinfo
  {author} {\bibfnamefont {A.}~\bibnamefont {Thamizhavel}}, \bibinfo {author}
  {\bibfnamefont {J.~F.}\ \bibnamefont {Steiner}}, \bibinfo {author}
  {\bibfnamefont {F.}~\bibnamefont {von Oppen}},\ and\ \bibinfo {author}
  {\bibfnamefont {M.~M.}\ \bibnamefont {Deshmukh}},\ }\bibfield  {title}
  {\bibinfo {title} {High-temperature {Josephson} diode},\ }\href
  {https://doi.org/10.1038/s41563-024-01804-4} {\bibfield  {journal} {\bibinfo
  {journal} {Nat. Mater.}\ }\textbf {\bibinfo {volume} {23}},\ \bibinfo {pages}
  {612} (\bibinfo {year} {2024})}\BibitemShut {NoStop}%
\bibitem [{\citenamefont {Le}\ \emph {et~al.}(2024)\citenamefont {Le},
  \citenamefont {Pan}, \citenamefont {Xu}, \citenamefont {Liu}, \citenamefont
  {Wang}, \citenamefont {Lou}, \citenamefont {Yang}, \citenamefont {Wang},
  \citenamefont {Yao}, \citenamefont {Wu},\ and\ \citenamefont
  {Lin}}]{Le_N_2024}%
  \BibitemOpen
  \bibfield  {author} {\bibinfo {author} {\bibfnamefont {T.}~\bibnamefont
  {Le}}, \bibinfo {author} {\bibfnamefont {Z.}~\bibnamefont {Pan}}, \bibinfo
  {author} {\bibfnamefont {Z.}~\bibnamefont {Xu}}, \bibinfo {author}
  {\bibfnamefont {J.}~\bibnamefont {Liu}}, \bibinfo {author} {\bibfnamefont
  {J.}~\bibnamefont {Wang}}, \bibinfo {author} {\bibfnamefont {Z.}~\bibnamefont
  {Lou}}, \bibinfo {author} {\bibfnamefont {X.}~\bibnamefont {Yang}}, \bibinfo
  {author} {\bibfnamefont {Z.}~\bibnamefont {Wang}}, \bibinfo {author}
  {\bibfnamefont {Y.}~\bibnamefont {Yao}}, \bibinfo {author} {\bibfnamefont
  {C.}~\bibnamefont {Wu}},\ and\ \bibinfo {author} {\bibfnamefont
  {X.}~\bibnamefont {Lin}},\ }\bibfield  {title} {\bibinfo {title}
  {Superconducting diode effect and interference patterns in kagome
  {CsV3Sb5}},\ }\href {https://doi.org/10.1038/s41586-024-07431-y} {\bibfield
  {journal} {\bibinfo  {journal} {Nature}\ }\textbf {\bibinfo {volume} {630}},\
  \bibinfo {pages} {64} (\bibinfo {year} {2024})}\BibitemShut {NoStop}%
\bibitem [{\citenamefont {Wan}\ \emph {et~al.}(2024)\citenamefont {Wan},
  \citenamefont {Qiu}, \citenamefont {Ren}, \citenamefont {Qian}, \citenamefont
  {Li}, \citenamefont {Xu}, \citenamefont {Zhou}, \citenamefont {Zhou},
  \citenamefont {Zhou}, \citenamefont {Wang}, \citenamefont {Yang},
  \citenamefont {Sofer}, \citenamefont {Huang}, \citenamefont {Wang},\ and\
  \citenamefont {Duan}}]{Wan_N_2024}%
  \BibitemOpen
  \bibfield  {author} {\bibinfo {author} {\bibfnamefont {Z.}~\bibnamefont
  {Wan}}, \bibinfo {author} {\bibfnamefont {G.}~\bibnamefont {Qiu}}, \bibinfo
  {author} {\bibfnamefont {H.}~\bibnamefont {Ren}}, \bibinfo {author}
  {\bibfnamefont {Q.}~\bibnamefont {Qian}}, \bibinfo {author} {\bibfnamefont
  {Y.}~\bibnamefont {Li}}, \bibinfo {author} {\bibfnamefont {D.}~\bibnamefont
  {Xu}}, \bibinfo {author} {\bibfnamefont {J.}~\bibnamefont {Zhou}}, \bibinfo
  {author} {\bibfnamefont {J.}~\bibnamefont {Zhou}}, \bibinfo {author}
  {\bibfnamefont {B.}~\bibnamefont {Zhou}}, \bibinfo {author} {\bibfnamefont
  {L.}~\bibnamefont {Wang}}, \bibinfo {author} {\bibfnamefont {T.-H.}\
  \bibnamefont {Yang}}, \bibinfo {author} {\bibfnamefont {Z.}~\bibnamefont
  {Sofer}}, \bibinfo {author} {\bibfnamefont {Y.}~\bibnamefont {Huang}},
  \bibinfo {author} {\bibfnamefont {K.~L.}\ \bibnamefont {Wang}},\ and\
  \bibinfo {author} {\bibfnamefont {X.}~\bibnamefont {Duan}},\ }\bibfield
  {title} {\bibinfo {title} {Unconventional superconductivity in chiral
  molecule–{TaS2} hybrid superlattices},\ }\href
  {https://doi.org/10.1038/s41586-024-07625-4} {\bibfield  {journal} {\bibinfo
  {journal} {Nature}\ }\textbf {\bibinfo {volume} {632}},\ \bibinfo {pages}
  {69} (\bibinfo {year} {2024})}\BibitemShut {NoStop}%
\bibitem [{\citenamefont {Liu}\ \emph {et~al.}(2024)\citenamefont {Liu},
  \citenamefont {Itahashi}, \citenamefont {Aoki}, \citenamefont {Dong},
  \citenamefont {Wang}, \citenamefont {Ogawa}, \citenamefont {Ideue},\ and\
  \citenamefont {Iwasa}}]{Liu_SA_2024}%
  \BibitemOpen
  \bibfield  {author} {\bibinfo {author} {\bibfnamefont {F.}~\bibnamefont
  {Liu}}, \bibinfo {author} {\bibfnamefont {Y.~M.}\ \bibnamefont {Itahashi}},
  \bibinfo {author} {\bibfnamefont {S.}~\bibnamefont {Aoki}}, \bibinfo {author}
  {\bibfnamefont {Y.}~\bibnamefont {Dong}}, \bibinfo {author} {\bibfnamefont
  {Z.}~\bibnamefont {Wang}}, \bibinfo {author} {\bibfnamefont {N.}~\bibnamefont
  {Ogawa}}, \bibinfo {author} {\bibfnamefont {T.}~\bibnamefont {Ideue}},\ and\
  \bibinfo {author} {\bibfnamefont {Y.}~\bibnamefont {Iwasa}},\ }\bibfield
  {title} {\bibinfo {title} {Superconducting diode effect under time-reversal
  symmetry},\ }\href {https://doi.org/10.1126/sciadv.ado1502} {\bibfield
  {journal} {\bibinfo  {journal} {Sci. Adv.}\ }\textbf {\bibinfo {volume}
  {10}},\ \bibinfo {pages} {eado1502} (\bibinfo {year} {2024})}\BibitemShut
  {NoStop}%
\bibitem [{\citenamefont {Qi}\ \emph {et~al.}(2025{\natexlab{a}})\citenamefont
  {Qi}, \citenamefont {Ge}, \citenamefont {Ji}, \citenamefont {Ai},
  \citenamefont {Ma}, \citenamefont {Wang}, \citenamefont {Cui}, \citenamefont
  {Liu}, \citenamefont {Wang},\ and\ \citenamefont {Wang}}]{Qi_NC_2025}%
  \BibitemOpen
  \bibfield  {author} {\bibinfo {author} {\bibfnamefont {S.}~\bibnamefont
  {Qi}}, \bibinfo {author} {\bibfnamefont {J.}~\bibnamefont {Ge}}, \bibinfo
  {author} {\bibfnamefont {C.}~\bibnamefont {Ji}}, \bibinfo {author}
  {\bibfnamefont {Y.}~\bibnamefont {Ai}}, \bibinfo {author} {\bibfnamefont
  {G.}~\bibnamefont {Ma}}, \bibinfo {author} {\bibfnamefont {Z.}~\bibnamefont
  {Wang}}, \bibinfo {author} {\bibfnamefont {Z.}~\bibnamefont {Cui}}, \bibinfo
  {author} {\bibfnamefont {Y.}~\bibnamefont {Liu}}, \bibinfo {author}
  {\bibfnamefont {Z.}~\bibnamefont {Wang}},\ and\ \bibinfo {author}
  {\bibfnamefont {J.}~\bibnamefont {Wang}},\ }\bibfield  {title} {\bibinfo
  {title} {High-temperature field-free superconducting diode effect in
  high-{Tc} cuprates},\ }\href {https://doi.org/10.1038/s41467-025-55880-4}
  {\bibfield  {journal} {\bibinfo  {journal} {Nat. Commun.}\ }\textbf {\bibinfo
  {volume} {16}},\ \bibinfo {pages} {531} (\bibinfo {year}
  {2025}{\natexlab{a}})}\BibitemShut {NoStop}%
\bibitem [{\citenamefont {Nagata}\ \emph {et~al.}(2025)\citenamefont {Nagata},
  \citenamefont {Aoki}, \citenamefont {Daido}, \citenamefont {Kasahara},
  \citenamefont {Kasahara}, \citenamefont {Ohshima}, \citenamefont {Ando},
  \citenamefont {Yanase}, \citenamefont {Matsuda},\ and\ \citenamefont
  {Shiraishi}}]{Nagata_PRL_2025}%
  \BibitemOpen
  \bibfield  {author} {\bibinfo {author} {\bibfnamefont {U.}~\bibnamefont
  {Nagata}}, \bibinfo {author} {\bibfnamefont {M.}~\bibnamefont {Aoki}},
  \bibinfo {author} {\bibfnamefont {A.}~\bibnamefont {Daido}}, \bibinfo
  {author} {\bibfnamefont {S.}~\bibnamefont {Kasahara}}, \bibinfo {author}
  {\bibfnamefont {Y.}~\bibnamefont {Kasahara}}, \bibinfo {author}
  {\bibfnamefont {R.}~\bibnamefont {Ohshima}}, \bibinfo {author} {\bibfnamefont
  {Y.}~\bibnamefont {Ando}}, \bibinfo {author} {\bibfnamefont {Y.}~\bibnamefont
  {Yanase}}, \bibinfo {author} {\bibfnamefont {Y.}~\bibnamefont {Matsuda}},\
  and\ \bibinfo {author} {\bibfnamefont {M.}~\bibnamefont {Shiraishi}},\
  }\bibfield  {title} {\bibinfo {title} {Field-{Free} {Superconducting} {Diode}
  {Effect} in {Layered} {Superconductor} {FeSe}},\ }\href
  {https://doi.org/10.1103/PhysRevLett.134.236703} {\bibfield  {journal}
  {\bibinfo  {journal} {Phys. Rev. Lett.}\ }\textbf {\bibinfo {volume} {134}},\
  \bibinfo {pages} {236703} (\bibinfo {year} {2025})}\BibitemShut {NoStop}%
\bibitem [{\citenamefont {Kudriashov}\ \emph {et~al.}(2025)\citenamefont
  {Kudriashov}, \citenamefont {Zhou}, \citenamefont {Hovhannisyan},
  \citenamefont {Frolov}, \citenamefont {Elesin}, \citenamefont {Wang},
  \citenamefont {Zharkova}, \citenamefont {Taniguchi}, \citenamefont
  {Watanabe}, \citenamefont {Liu}, \citenamefont {Novoselov}, \citenamefont
  {Yashina}, \citenamefont {Zhou},\ and\ \citenamefont
  {Bandurin}}]{Kudriashov_SA_2025}%
  \BibitemOpen
  \bibfield  {author} {\bibinfo {author} {\bibfnamefont {A.}~\bibnamefont
  {Kudriashov}}, \bibinfo {author} {\bibfnamefont {X.}~\bibnamefont {Zhou}},
  \bibinfo {author} {\bibfnamefont {R.~A.}\ \bibnamefont {Hovhannisyan}},
  \bibinfo {author} {\bibfnamefont {A.~S.}\ \bibnamefont {Frolov}}, \bibinfo
  {author} {\bibfnamefont {L.}~\bibnamefont {Elesin}}, \bibinfo {author}
  {\bibfnamefont {Y.~B.}\ \bibnamefont {Wang}}, \bibinfo {author}
  {\bibfnamefont {E.~V.}\ \bibnamefont {Zharkova}}, \bibinfo {author}
  {\bibfnamefont {T.}~\bibnamefont {Taniguchi}}, \bibinfo {author}
  {\bibfnamefont {K.}~\bibnamefont {Watanabe}}, \bibinfo {author}
  {\bibfnamefont {Z.}~\bibnamefont {Liu}}, \bibinfo {author} {\bibfnamefont
  {K.~S.}\ \bibnamefont {Novoselov}}, \bibinfo {author} {\bibfnamefont {L.~V.}\
  \bibnamefont {Yashina}}, \bibinfo {author} {\bibfnamefont {X.}~\bibnamefont
  {Zhou}},\ and\ \bibinfo {author} {\bibfnamefont {D.~A.}\ \bibnamefont
  {Bandurin}},\ }\bibfield  {title} {\bibinfo {title} {Non-{Majorana} origin of
  anomalous current-phase relation and {Josephson} diode effect in
  {Bi2Se3}/{NbSe2} {Josephson} junctions},\ }\href
  {https://doi.org/10.1126/sciadv.adw6925} {\bibfield  {journal} {\bibinfo
  {journal} {Sci. Adv.}\ }\textbf {\bibinfo {volume} {11}},\ \bibinfo {pages}
  {eadw6925} (\bibinfo {year} {2025})}\BibitemShut {NoStop}%
\bibitem [{\citenamefont {Shi}\ \emph {et~al.}(2025)\citenamefont {Shi},
  \citenamefont {Dou}, \citenamefont {Pan}, \citenamefont {Li}, \citenamefont
  {Li}, \citenamefont {Wang}, \citenamefont {Zhang}, \citenamefont {Guo},
  \citenamefont {Deng}, \citenamefont {Tong}, \citenamefont {Lyu},
  \citenamefont {Li}, \citenamefont {Qu}, \citenamefont {Liu}, \citenamefont
  {Zhao}, \citenamefont {Hu}, \citenamefont {Lu},\ and\ \citenamefont
  {Shen}}]{Shi_a_2025}%
  \BibitemOpen
  \bibfield  {author} {\bibinfo {author} {\bibfnamefont {X.}~\bibnamefont
  {Shi}}, \bibinfo {author} {\bibfnamefont {Z.}~\bibnamefont {Dou}}, \bibinfo
  {author} {\bibfnamefont {D.}~\bibnamefont {Pan}}, \bibinfo {author}
  {\bibfnamefont {G.}~\bibnamefont {Li}}, \bibinfo {author} {\bibfnamefont
  {Y.}~\bibnamefont {Li}}, \bibinfo {author} {\bibfnamefont {A.}~\bibnamefont
  {Wang}}, \bibinfo {author} {\bibfnamefont {Z.}~\bibnamefont {Zhang}},
  \bibinfo {author} {\bibfnamefont {X.}~\bibnamefont {Guo}}, \bibinfo {author}
  {\bibfnamefont {X.}~\bibnamefont {Deng}}, \bibinfo {author} {\bibfnamefont
  {B.}~\bibnamefont {Tong}}, \bibinfo {author} {\bibfnamefont {Z.}~\bibnamefont
  {Lyu}}, \bibinfo {author} {\bibfnamefont {P.}~\bibnamefont {Li}}, \bibinfo
  {author} {\bibfnamefont {F.}~\bibnamefont {Qu}}, \bibinfo {author}
  {\bibfnamefont {G.}~\bibnamefont {Liu}}, \bibinfo {author} {\bibfnamefont
  {J.}~\bibnamefont {Zhao}}, \bibinfo {author} {\bibfnamefont {J.}~\bibnamefont
  {Hu}}, \bibinfo {author} {\bibfnamefont {L.}~\bibnamefont {Lu}},\ and\
  \bibinfo {author} {\bibfnamefont {J.}~\bibnamefont {Shen}},\ }\bibfield
  {title} {\bibinfo {title} {Circuit-level-configurable {Zero}-field
  {Superconducting} {Diodes}: {A} {Universal} {Platform} {Beyond} {Intrinsic}
  {Symmetry} {Breaking}},\ }\href {http://arxiv.org/abs/2505.18330} {\bibfield
  {journal} {\bibinfo  {journal} {arXiv:2505.18330}\ } (\bibinfo {year}
  {2025})}\BibitemShut {NoStop}%
\bibitem [{\citenamefont {Wang}\ \emph
  {et~al.}(2025{\natexlab{a}})\citenamefont {Wang}, \citenamefont {Zhu},
  \citenamefont {Bai}, \citenamefont {Lyu}, \citenamefont {Yang}, \citenamefont
  {Zhao}, \citenamefont {Zhou}, \citenamefont {Xue},\ and\ \citenamefont
  {Zhang}}]{Wang_NP_2025}%
  \BibitemOpen
  \bibfield  {author} {\bibinfo {author} {\bibfnamefont {H.}~\bibnamefont
  {Wang}}, \bibinfo {author} {\bibfnamefont {Y.}~\bibnamefont {Zhu}}, \bibinfo
  {author} {\bibfnamefont {Z.}~\bibnamefont {Bai}}, \bibinfo {author}
  {\bibfnamefont {Z.}~\bibnamefont {Lyu}}, \bibinfo {author} {\bibfnamefont
  {J.}~\bibnamefont {Yang}}, \bibinfo {author} {\bibfnamefont {L.}~\bibnamefont
  {Zhao}}, \bibinfo {author} {\bibfnamefont {X.~J.}\ \bibnamefont {Zhou}},
  \bibinfo {author} {\bibfnamefont {Q.-K.}\ \bibnamefont {Xue}},\ and\ \bibinfo
  {author} {\bibfnamefont {D.}~\bibnamefont {Zhang}},\ }\bibfield  {title}
  {\bibinfo {title} {Quantum superconducting diode effect with perfect
  efficiency above liquid-nitrogen temperature},\ }\href
  {https://doi.org/10.1038/s41567-025-03098-y} {\bibfield  {journal} {\bibinfo
  {journal} {Nat. Phys.}\ }\textbf {\bibinfo {volume} {22}},\ \bibinfo {pages}
  {47} (\bibinfo {year} {2025}{\natexlab{a}})}\BibitemShut {NoStop}%
\bibitem [{\citenamefont {Nagahama}\ \emph {et~al.}(2025)\citenamefont
  {Nagahama}, \citenamefont {Sato}, \citenamefont {Kawamura}, \citenamefont
  {Belopolski}, \citenamefont {Yoshimi}, \citenamefont {Tsukazaki},
  \citenamefont {Kanazawa}, \citenamefont {Takahashi}, \citenamefont
  {Kawasaki},\ and\ \citenamefont {Tokura}}]{Nagahama_PRL_2025}%
  \BibitemOpen
  \bibfield  {author} {\bibinfo {author} {\bibfnamefont {S.}~\bibnamefont
  {Nagahama}}, \bibinfo {author} {\bibfnamefont {Y.}~\bibnamefont {Sato}},
  \bibinfo {author} {\bibfnamefont {M.}~\bibnamefont {Kawamura}}, \bibinfo
  {author} {\bibfnamefont {I.}~\bibnamefont {Belopolski}}, \bibinfo {author}
  {\bibfnamefont {R.}~\bibnamefont {Yoshimi}}, \bibinfo {author} {\bibfnamefont
  {A.}~\bibnamefont {Tsukazaki}}, \bibinfo {author} {\bibfnamefont
  {N.}~\bibnamefont {Kanazawa}}, \bibinfo {author} {\bibfnamefont {K.~S.}\
  \bibnamefont {Takahashi}}, \bibinfo {author} {\bibfnamefont {M.}~\bibnamefont
  {Kawasaki}},\ and\ \bibinfo {author} {\bibfnamefont {Y.}~\bibnamefont
  {Tokura}},\ }\bibfield  {title} {\bibinfo {title} {Two-dimensional
  superconducting diode effect in topological insulator/superconductor
  heterostructure},\ }\bibfield  {journal} {\bibinfo  {journal} {Phys. Rev.
  Lett.}\ }\textbf {\bibinfo {volume} {135}},\ \href
  {https://doi.org/10.1103/tyxq-t1bc} {10.1103/tyxq-t1bc} (\bibinfo {year}
  {2025})\BibitemShut {NoStop}%
\bibitem [{\citenamefont {Zhang}\ \emph {et~al.}(2026)\citenamefont {Zhang},
  \citenamefont {Wang}, \citenamefont {Wang}, \citenamefont {Pan},
  \citenamefont {Wang}, \citenamefont {Cai}, \citenamefont {Liu}, \citenamefont
  {Qu}, \citenamefont {Zhu}, \citenamefont {Ning}, \citenamefont {Xi},
  \citenamefont {Zhang}, \citenamefont {Hao}, \citenamefont {Zheng},\ and\
  \citenamefont {Tian}}]{Zhang_PRX_2026}%
  \BibitemOpen
  \bibfield  {author} {\bibinfo {author} {\bibfnamefont {H.}~\bibnamefont
  {Zhang}}, \bibinfo {author} {\bibfnamefont {C.}~\bibnamefont {Wang}},
  \bibinfo {author} {\bibfnamefont {R.}~\bibnamefont {Wang}}, \bibinfo {author}
  {\bibfnamefont {S.}~\bibnamefont {Pan}}, \bibinfo {author} {\bibfnamefont
  {H.}~\bibnamefont {Wang}}, \bibinfo {author} {\bibfnamefont {J.}~\bibnamefont
  {Cai}}, \bibinfo {author} {\bibfnamefont {Y.}~\bibnamefont {Liu}}, \bibinfo
  {author} {\bibfnamefont {Z.}~\bibnamefont {Qu}}, \bibinfo {author}
  {\bibfnamefont {X.}~\bibnamefont {Zhu}}, \bibinfo {author} {\bibfnamefont
  {W.}~\bibnamefont {Ning}}, \bibinfo {author} {\bibfnamefont {C.}~\bibnamefont
  {Xi}}, \bibinfo {author} {\bibfnamefont {J.}~\bibnamefont {Zhang}}, \bibinfo
  {author} {\bibfnamefont {N.}~\bibnamefont {Hao}}, \bibinfo {author}
  {\bibfnamefont {G.}~\bibnamefont {Zheng}},\ and\ \bibinfo {author}
  {\bibfnamefont {M.}~\bibnamefont {Tian}},\ }\bibfield  {title} {\bibinfo
  {title} {General framework enabling polarity-tunable time-reversal symmetric
  superconducting diode effects in gate-defined homojunctions},\ }\href
  {https://doi.org/10.1103/wm2k-vlvc} {\bibfield  {journal} {\bibinfo
  {journal} {Phys. Rev. X}\ }\textbf {\bibinfo {volume} {16}},\ \bibinfo
  {pages} {021046} (\bibinfo {year} {2026})}\BibitemShut {NoStop}%
\bibitem [{\citenamefont {Chen}\ \emph {et~al.}(2026)\citenamefont {Chen},
  \citenamefont {Lou}, \citenamefont {Ye}, \citenamefont {Tan}, \citenamefont
  {Wang}, \citenamefont {Zhang}, \citenamefont {Wang}, \citenamefont {Liao},
  \citenamefont {Zhang}, \citenamefont {Liao},\ and\ \citenamefont
  {Yu}}]{Chen_PRL_2026}%
  \BibitemOpen
  \bibfield  {author} {\bibinfo {author} {\bibfnamefont {J.-J.}\ \bibnamefont
  {Chen}}, \bibinfo {author} {\bibfnamefont {H.-X.}\ \bibnamefont {Lou}},
  \bibinfo {author} {\bibfnamefont {X.-G.}\ \bibnamefont {Ye}}, \bibinfo
  {author} {\bibfnamefont {Z.-B.}\ \bibnamefont {Tan}}, \bibinfo {author}
  {\bibfnamefont {A.-Q.}\ \bibnamefont {Wang}}, \bibinfo {author}
  {\bibfnamefont {Z.-T.}\ \bibnamefont {Zhang}}, \bibinfo {author}
  {\bibfnamefont {Y.-X.}\ \bibnamefont {Wang}}, \bibinfo {author}
  {\bibfnamefont {X.}~\bibnamefont {Liao}}, \bibinfo {author} {\bibfnamefont
  {Z.-S.}\ \bibnamefont {Zhang}}, \bibinfo {author} {\bibfnamefont {Z.-M.}\
  \bibnamefont {Liao}},\ and\ \bibinfo {author} {\bibfnamefont {D.-P.}\
  \bibnamefont {Yu}},\ }\bibfield  {title} {\bibinfo {title} {Magnetic-flux
  tuning of second harmonic content in intrinsic
  ${\mathrm{csv}}_{3}{\mathrm{sb}}_{5}$ josephson junction},\ }\href
  {https://doi.org/10.1103/34th-gqsy} {\bibfield  {journal} {\bibinfo
  {journal} {Phys. Rev. Lett.}\ }\textbf {\bibinfo {volume} {136}},\ \bibinfo
  {pages} {096301} (\bibinfo {year} {2026})}\BibitemShut {NoStop}%
\bibitem [{\citenamefont {Mangold}\ \emph {et~al.}(2026)\citenamefont
  {Mangold}, \citenamefont {Bauriedl}, \citenamefont {Berger}, \citenamefont
  {Yu-Cheng}, \citenamefont {Meier}, \citenamefont {Kronseder}, \citenamefont
  {Hakonen}, \citenamefont {Back}, \citenamefont {Strunk},\ and\ \citenamefont
  {Suri}}]{Mangold_PRL_2026}%
  \BibitemOpen
  \bibfield  {author} {\bibinfo {author} {\bibfnamefont {M.}~\bibnamefont
  {Mangold}}, \bibinfo {author} {\bibfnamefont {L.}~\bibnamefont {Bauriedl}},
  \bibinfo {author} {\bibfnamefont {J.}~\bibnamefont {Berger}}, \bibinfo
  {author} {\bibfnamefont {C.}~\bibnamefont {Yu-Cheng}}, \bibinfo {author}
  {\bibfnamefont {T.~N.~G.}\ \bibnamefont {Meier}}, \bibinfo {author}
  {\bibfnamefont {M.}~\bibnamefont {Kronseder}}, \bibinfo {author}
  {\bibfnamefont {P.}~\bibnamefont {Hakonen}}, \bibinfo {author} {\bibfnamefont
  {C.~H.}\ \bibnamefont {Back}}, \bibinfo {author} {\bibfnamefont
  {C.}~\bibnamefont {Strunk}},\ and\ \bibinfo {author} {\bibfnamefont
  {D.}~\bibnamefont {Suri}},\ }\bibfield  {title} {\bibinfo {title}
  {Magnetochiral anisotropy in josephson diode effect of all-metallic lateral
  junctions with interfacial rashba spin-orbit coupling},\ }\href
  {https://doi.org/10.1103/n9g9-99wl} {\bibfield  {journal} {\bibinfo
  {journal} {Phys. Rev. Lett.}\ }\textbf {\bibinfo {volume} {136}},\ \bibinfo
  {pages} {206002} (\bibinfo {year} {2026})}\BibitemShut {NoStop}%
\bibitem [{\citenamefont {Zhu}\ \emph {et~al.}(2026)\citenamefont {Zhu},
  \citenamefont {Le}, \citenamefont {Pei}, \citenamefont {Li}, \citenamefont
  {Liao}, \citenamefont {Zhao}, \citenamefont {Zhao}, \citenamefont {Wang},
  \citenamefont {Wu}, \citenamefont {Zhang}, \citenamefont {Wu}, \citenamefont
  {Fu}, \citenamefont {Lü}, \citenamefont {Yang}, \citenamefont {Shen},
  \citenamefont {Li}, \citenamefont {Chen}, \citenamefont {Lin}, \citenamefont
  {He},\ and\ \citenamefont {Qi}}]{Zhu_PRL_2026}%
  \BibitemOpen
  \bibfield  {author} {\bibinfo {author} {\bibfnamefont {S.}~\bibnamefont
  {Zhu}}, \bibinfo {author} {\bibfnamefont {T.}~\bibnamefont {Le}}, \bibinfo
  {author} {\bibfnamefont {C.}~\bibnamefont {Pei}}, \bibinfo {author}
  {\bibfnamefont {C.}~\bibnamefont {Li}}, \bibinfo {author} {\bibfnamefont
  {Y.}~\bibnamefont {Liao}}, \bibinfo {author} {\bibfnamefont {Y.}~\bibnamefont
  {Zhao}}, \bibinfo {author} {\bibfnamefont {L.}~\bibnamefont {Zhao}}, \bibinfo
  {author} {\bibfnamefont {Q.}~\bibnamefont {Wang}}, \bibinfo {author}
  {\bibfnamefont {J.}~\bibnamefont {Wu}}, \bibinfo {author} {\bibfnamefont
  {Q.}~\bibnamefont {Zhang}}, \bibinfo {author} {\bibfnamefont
  {Y.}~\bibnamefont {Wu}}, \bibinfo {author} {\bibfnamefont {T.}~\bibnamefont
  {Fu}}, \bibinfo {author} {\bibfnamefont {X.}~\bibnamefont {Lü}}, \bibinfo
  {author} {\bibfnamefont {W.}~\bibnamefont {Yang}}, \bibinfo {author}
  {\bibfnamefont {J.}~\bibnamefont {Shen}}, \bibinfo {author} {\bibfnamefont
  {J.}~\bibnamefont {Li}}, \bibinfo {author} {\bibfnamefont {Y.}~\bibnamefont
  {Chen}}, \bibinfo {author} {\bibfnamefont {X.}~\bibnamefont {Lin}}, \bibinfo
  {author} {\bibfnamefont {W.-Y.}\ \bibnamefont {He}},\ and\ \bibinfo {author}
  {\bibfnamefont {Y.}~\bibnamefont {Qi}},\ }\bibfield  {title} {\bibinfo
  {title} {Pressure-induced magnetic-field-free superconducting diode effect in
  nbse 2 flake},\ }\bibfield  {journal} {\bibinfo  {journal} {Phys. Rev.
  Lett.}\ }\textbf {\bibinfo {volume} {137}},\ \href
  {https://doi.org/10.1103/8wnw-16qt} {10.1103/8wnw-16qt} (\bibinfo {year}
  {2026})\BibitemShut {NoStop}%
\bibitem [{\citenamefont {Wei}\ \emph {et~al.}(2026)\citenamefont {Wei},
  \citenamefont {Qiao}, \citenamefont {Lyu}, \citenamefont {Wang},
  \citenamefont {Li}, \citenamefont {Cadorim}, \citenamefont {Zhang},
  \citenamefont {Yue}, \citenamefont {Li}, \citenamefont {Song}, \citenamefont
  {Wang}, \citenamefont {Wang}, \citenamefont {Milošević}, \citenamefont
  {Wang}, \citenamefont {Wang},\ and\ \citenamefont {Wu}}]{Wei_NSR_2026}%
  \BibitemOpen
  \bibfield  {author} {\bibinfo {author} {\bibfnamefont {Z.}~\bibnamefont
  {Wei}}, \bibinfo {author} {\bibfnamefont {Y.}~\bibnamefont {Qiao}}, \bibinfo
  {author} {\bibfnamefont {Y.-Y.}\ \bibnamefont {Lyu}}, \bibinfo {author}
  {\bibfnamefont {D.}~\bibnamefont {Wang}}, \bibinfo {author} {\bibfnamefont
  {T.}~\bibnamefont {Li}}, \bibinfo {author} {\bibfnamefont {L.~R.}\
  \bibnamefont {Cadorim}}, \bibinfo {author} {\bibfnamefont {P.}~\bibnamefont
  {Zhang}}, \bibinfo {author} {\bibfnamefont {W.-C.}\ \bibnamefont {Yue}},
  \bibinfo {author} {\bibfnamefont {D.}~\bibnamefont {Li}}, \bibinfo {author}
  {\bibfnamefont {Z.}~\bibnamefont {Song}}, \bibinfo {author} {\bibfnamefont
  {Z.}~\bibnamefont {Wang}}, \bibinfo {author} {\bibfnamefont {Y.}~\bibnamefont
  {Wang}}, \bibinfo {author} {\bibfnamefont {M.~V.}\ \bibnamefont
  {Milošević}}, \bibinfo {author} {\bibfnamefont {Y.-L.}\ \bibnamefont
  {Wang}}, \bibinfo {author} {\bibfnamefont {H.}~\bibnamefont {Wang}},\ and\
  \bibinfo {author} {\bibfnamefont {P.}~\bibnamefont {Wu}},\ }\bibfield
  {title} {\bibinfo {title} {Scalable high-temperature superconducting diodes
  enabled by intrinsic josephson junctions},\ }\bibfield  {journal} {\bibinfo
  {journal} {Natl. Sci. Rev.}\ }\href {https://doi.org/10.1093/nsr/nwag285}
  {10.1093/nsr/nwag285} (\bibinfo {year} {2026})\BibitemShut {NoStop}%
\bibitem [{\citenamefont {Misaki}\ and\ \citenamefont
  {Nagaosa}(2021)}]{Misaki_PRB_2021}%
  \BibitemOpen
  \bibfield  {author} {\bibinfo {author} {\bibfnamefont {K.}~\bibnamefont
  {Misaki}}\ and\ \bibinfo {author} {\bibfnamefont {N.}~\bibnamefont
  {Nagaosa}},\ }\bibfield  {title} {\bibinfo {title} {Theory of the
  nonreciprocal {{Josephson}} effect},\ }\href
  {https://doi.org/10.1103/PhysRevB.103.245302} {\bibfield  {journal} {\bibinfo
   {journal} {Phys. Rev. B}\ }\textbf {\bibinfo {volume} {103}},\ \bibinfo
  {pages} {245302} (\bibinfo {year} {2021})}\BibitemShut {NoStop}%
\bibitem [{\citenamefont {Scammell}\ \emph {et~al.}(2022)\citenamefont
  {Scammell}, \citenamefont {Li},\ and\ \citenamefont
  {Scheurer}}]{Scammell_2M_2022}%
  \BibitemOpen
  \bibfield  {author} {\bibinfo {author} {\bibfnamefont {H.~D.}\ \bibnamefont
  {Scammell}}, \bibinfo {author} {\bibfnamefont {J.~I.~A.}\ \bibnamefont
  {Li}},\ and\ \bibinfo {author} {\bibfnamefont {M.~S.}\ \bibnamefont
  {Scheurer}},\ }\bibfield  {title} {\bibinfo {title} {Theory of zero-field
  superconducting diode effect in twisted trilayer graphene},\ }\href
  {https://doi.org/10.1088/2053-1583/ac5b16} {\bibfield  {journal} {\bibinfo
  {journal} {2D Mater.}\ }\textbf {\bibinfo {volume} {9}},\ \bibinfo {pages}
  {025027} (\bibinfo {year} {2022})}\BibitemShut {NoStop}%
\bibitem [{\citenamefont {Wang}\ \emph
  {et~al.}(2025{\natexlab{b}})\citenamefont {Wang}, \citenamefont {Wang},\ and\
  \citenamefont {Wu}}]{Wang_a_2022}%
  \BibitemOpen
  \bibfield  {author} {\bibinfo {author} {\bibfnamefont {D.}~\bibnamefont
  {Wang}}, \bibinfo {author} {\bibfnamefont {Q.-H.}\ \bibnamefont {Wang}},\
  and\ \bibinfo {author} {\bibfnamefont {C.}~\bibnamefont {Wu}},\ }\bibfield
  {title} {\bibinfo {title} {Current-reversion symmetry breaking and the {DC}
  {Josephson} diode effect},\ }\href
  {https://doi.org/10.1016/j.scib.2025.11.011} {\bibfield  {journal} {\bibinfo
  {journal} {Sci. Bull.}\ }\textbf {\bibinfo {volume} {70}},\ \bibinfo {pages}
  {4181} (\bibinfo {year} {2025}{\natexlab{b}})}\BibitemShut {NoStop}%
\bibitem [{\citenamefont {He}\ \emph {et~al.}(2022)\citenamefont {He},
  \citenamefont {Tanaka},\ and\ \citenamefont {Nagaosa}}]{He_NJP_2022}%
  \BibitemOpen
  \bibfield  {author} {\bibinfo {author} {\bibfnamefont {J.~J.}\ \bibnamefont
  {He}}, \bibinfo {author} {\bibfnamefont {Y.}~\bibnamefont {Tanaka}},\ and\
  \bibinfo {author} {\bibfnamefont {N.}~\bibnamefont {Nagaosa}},\ }\bibfield
  {title} {\bibinfo {title} {A phenomenological theory of superconductor
  diodes},\ }\href {https://doi.org/10.1088/1367-2630/ac6766} {\bibfield
  {journal} {\bibinfo  {journal} {New J. Phys.}\ }\textbf {\bibinfo {volume}
  {24}},\ \bibinfo {pages} {053014} (\bibinfo {year} {2022})}\BibitemShut
  {NoStop}%
\bibitem [{\citenamefont {Legg}\ \emph {et~al.}(2022)\citenamefont {Legg},
  \citenamefont {Loss},\ and\ \citenamefont {Klinovaja}}]{Legg_PRB_2022}%
  \BibitemOpen
  \bibfield  {author} {\bibinfo {author} {\bibfnamefont {H.~F.}\ \bibnamefont
  {Legg}}, \bibinfo {author} {\bibfnamefont {D.}~\bibnamefont {Loss}},\ and\
  \bibinfo {author} {\bibfnamefont {J.}~\bibnamefont {Klinovaja}},\ }\bibfield
  {title} {\bibinfo {title} {Superconducting diode effect due to magnetochiral
  anisotropy in topological insulators and rashba nanowires},\ }\href
  {https://doi.org/10.1103/PhysRevB.106.104501} {\bibfield  {journal} {\bibinfo
   {journal} {Phys. Rev. B}\ }\textbf {\bibinfo {volume} {106}},\ \bibinfo
  {pages} {104501} (\bibinfo {year} {2022})}\BibitemShut {NoStop}%
\bibitem [{\citenamefont {Fominov}\ and\ \citenamefont
  {Mikhailov}(2022)}]{Fominov_PRB_2022}%
  \BibitemOpen
  \bibfield  {author} {\bibinfo {author} {\bibfnamefont {Y.~V.}\ \bibnamefont
  {Fominov}}\ and\ \bibinfo {author} {\bibfnamefont {D.~S.}\ \bibnamefont
  {Mikhailov}},\ }\bibfield  {title} {\bibinfo {title} {Asymmetric
  higher-harmonic squid as a josephson diode},\ }\href
  {https://doi.org/10.1103/PhysRevB.106.134514} {\bibfield  {journal} {\bibinfo
   {journal} {Phys. Rev. B}\ }\textbf {\bibinfo {volume} {106}},\ \bibinfo
  {pages} {134514} (\bibinfo {year} {2022})}\BibitemShut {NoStop}%
\bibitem [{\citenamefont {Kokkeler}\ \emph {et~al.}(2022)\citenamefont
  {Kokkeler}, \citenamefont {Golubov},\ and\ \citenamefont
  {Bergeret}}]{Kokkeler_PRB_2022}%
  \BibitemOpen
  \bibfield  {author} {\bibinfo {author} {\bibfnamefont {T.~H.}\ \bibnamefont
  {Kokkeler}}, \bibinfo {author} {\bibfnamefont {A.~A.}\ \bibnamefont
  {Golubov}},\ and\ \bibinfo {author} {\bibfnamefont {F.~S.}\ \bibnamefont
  {Bergeret}},\ }\bibfield  {title} {\bibinfo {title} {Field-free anomalous
  junction and superconducting diode effect in spin-split
  superconductor/topological insulator junctions},\ }\href
  {https://doi.org/10.1103/PhysRevB.106.214504} {\bibfield  {journal} {\bibinfo
   {journal} {Phys. Rev. B}\ }\textbf {\bibinfo {volume} {106}},\ \bibinfo
  {pages} {214504} (\bibinfo {year} {2022})}\BibitemShut {NoStop}%
\bibitem [{\citenamefont {Daido}\ \emph {et~al.}(2022)\citenamefont {Daido},
  \citenamefont {Ikeda},\ and\ \citenamefont {Yanase}}]{Daido_PRL_2022}%
  \BibitemOpen
  \bibfield  {author} {\bibinfo {author} {\bibfnamefont {A.}~\bibnamefont
  {Daido}}, \bibinfo {author} {\bibfnamefont {Y.}~\bibnamefont {Ikeda}},\ and\
  \bibinfo {author} {\bibfnamefont {Y.}~\bibnamefont {Yanase}},\ }\bibfield
  {title} {\bibinfo {title} {Intrinsic {{Superconducting Diode Effect}}},\
  }\href {https://doi.org/10.1103/PhysRevLett.128.037001} {\bibfield  {journal}
  {\bibinfo  {journal} {Phys. Rev. Lett.}\ }\textbf {\bibinfo {volume} {128}},\
  \bibinfo {pages} {037001} (\bibinfo {year} {2022})}\BibitemShut {NoStop}%
\bibitem [{\citenamefont {Souto}\ \emph {et~al.}(2022)\citenamefont {Souto},
  \citenamefont {Leijnse},\ and\ \citenamefont {Schrade}}]{Souto_PRL_2022}%
  \BibitemOpen
  \bibfield  {author} {\bibinfo {author} {\bibfnamefont {R.~S.}\ \bibnamefont
  {Souto}}, \bibinfo {author} {\bibfnamefont {M.}~\bibnamefont {Leijnse}},\
  and\ \bibinfo {author} {\bibfnamefont {C.}~\bibnamefont {Schrade}},\
  }\bibfield  {title} {\bibinfo {title} {Josephson {Diode} {Effect} in
  {Supercurrent} {Interferometers}},\ }\href
  {https://doi.org/10.1103/PhysRevLett.129.267702} {\bibfield  {journal}
  {\bibinfo  {journal} {Phys. Rev. Lett.}\ }\textbf {\bibinfo {volume} {129}},\
  \bibinfo {pages} {267702} (\bibinfo {year} {2022})}\BibitemShut {NoStop}%
\bibitem [{\citenamefont {Zinkl}\ \emph {et~al.}(2022)\citenamefont {Zinkl},
  \citenamefont {Hamamoto},\ and\ \citenamefont {Sigrist}}]{Zinkl_PRR_2022}%
  \BibitemOpen
  \bibfield  {author} {\bibinfo {author} {\bibfnamefont {B.}~\bibnamefont
  {Zinkl}}, \bibinfo {author} {\bibfnamefont {K.}~\bibnamefont {Hamamoto}},\
  and\ \bibinfo {author} {\bibfnamefont {M.}~\bibnamefont {Sigrist}},\
  }\bibfield  {title} {\bibinfo {title} {Symmetry conditions for the
  superconducting diode effect in chiral superconductors},\ }\href
  {https://doi.org/10.1103/PhysRevResearch.4.033167} {\bibfield  {journal}
  {\bibinfo  {journal} {Phys. Rev. Res.}\ }\textbf {\bibinfo {volume} {4}},\
  \bibinfo {pages} {033167} (\bibinfo {year} {2022})}\BibitemShut {NoStop}%
\bibitem [{\citenamefont {Zhang}\ \emph {et~al.}(2022)\citenamefont {Zhang},
  \citenamefont {Gu}, \citenamefont {Li}, \citenamefont {Hu},\ and\
  \citenamefont {Jiang}}]{Zhang_PRX_2022}%
  \BibitemOpen
  \bibfield  {author} {\bibinfo {author} {\bibfnamefont {Y.}~\bibnamefont
  {Zhang}}, \bibinfo {author} {\bibfnamefont {Y.}~\bibnamefont {Gu}}, \bibinfo
  {author} {\bibfnamefont {P.}~\bibnamefont {Li}}, \bibinfo {author}
  {\bibfnamefont {J.}~\bibnamefont {Hu}},\ and\ \bibinfo {author}
  {\bibfnamefont {K.}~\bibnamefont {Jiang}},\ }\bibfield  {title} {\bibinfo
  {title} {General {Theory} of {Josephson} {Diodes}},\ }\href
  {https://doi.org/10.1103/PhysRevX.12.041013} {\bibfield  {journal} {\bibinfo
  {journal} {Phys. Rev. X}\ }\textbf {\bibinfo {volume} {12}},\ \bibinfo
  {pages} {041013} (\bibinfo {year} {2022})}\BibitemShut {NoStop}%
\bibitem [{\citenamefont {Yuan}\ and\ \citenamefont
  {Fu}(2022)}]{Yuan_PotNAoS_2022}%
  \BibitemOpen
  \bibfield  {author} {\bibinfo {author} {\bibfnamefont {N.~F.~Q.}\
  \bibnamefont {Yuan}}\ and\ \bibinfo {author} {\bibfnamefont {L.}~\bibnamefont
  {Fu}},\ }\bibfield  {title} {\bibinfo {title} {Supercurrent diode effect and
  finite-momentum superconductors},\ }\href
  {https://doi.org/10.1073/pnas.2119548119} {\bibfield  {journal} {\bibinfo
  {journal} {Proc. Natl. Acad. Sci.}\ }\textbf {\bibinfo {volume} {119}},\
  \bibinfo {pages} {e2119548119} (\bibinfo {year} {2022})}\BibitemShut
  {NoStop}%
\bibitem [{\citenamefont {Davydova}\ \emph {et~al.}(2022)\citenamefont
  {Davydova}, \citenamefont {Prembabu},\ and\ \citenamefont
  {Fu}}]{Davydova_SA_2022}%
  \BibitemOpen
  \bibfield  {author} {\bibinfo {author} {\bibfnamefont {M.}~\bibnamefont
  {Davydova}}, \bibinfo {author} {\bibfnamefont {S.}~\bibnamefont {Prembabu}},\
  and\ \bibinfo {author} {\bibfnamefont {L.}~\bibnamefont {Fu}},\ }\bibfield
  {title} {\bibinfo {title} {Universal {Josephson} diode effect},\ }\href
  {https://doi.org/10.1126/sciadv.abo0309} {\bibfield  {journal} {\bibinfo
  {journal} {Sci. Adv.}\ }\textbf {\bibinfo {volume} {8}},\ \bibinfo {pages}
  {eabo0309} (\bibinfo {year} {2022})}\BibitemShut {NoStop}%
\bibitem [{\citenamefont {He}\ \emph {et~al.}(2023)\citenamefont {He},
  \citenamefont {Tanaka},\ and\ \citenamefont {Nagaosa}}]{He_NC_2023}%
  \BibitemOpen
  \bibfield  {author} {\bibinfo {author} {\bibfnamefont {J.~J.}\ \bibnamefont
  {He}}, \bibinfo {author} {\bibfnamefont {Y.}~\bibnamefont {Tanaka}},\ and\
  \bibinfo {author} {\bibfnamefont {N.}~\bibnamefont {Nagaosa}},\ }\bibfield
  {title} {\bibinfo {title} {The supercurrent diode effect and nonreciprocal
  paraconductivity due to the chiral structure of nanotubes},\ }\href
  {https://doi.org/10.1038/s41467-023-39083-3} {\bibfield  {journal} {\bibinfo
  {journal} {Nat. Commun.}\ }\textbf {\bibinfo {volume} {14}},\ \bibinfo
  {pages} {3330} (\bibinfo {year} {2023})}\BibitemShut {NoStop}%
\bibitem [{\citenamefont {Hu}\ \emph {et~al.}(2023)\citenamefont {Hu},
  \citenamefont {Sun}, \citenamefont {Xie},\ and\ \citenamefont
  {Law}}]{Hu_PRL_2023}%
  \BibitemOpen
  \bibfield  {author} {\bibinfo {author} {\bibfnamefont {J.-X.}\ \bibnamefont
  {Hu}}, \bibinfo {author} {\bibfnamefont {Z.-T.}\ \bibnamefont {Sun}},
  \bibinfo {author} {\bibfnamefont {Y.-M.}\ \bibnamefont {Xie}},\ and\ \bibinfo
  {author} {\bibfnamefont {K.}~\bibnamefont {Law}},\ }\bibfield  {title}
  {\bibinfo {title} {Josephson {Diode} {Effect} {Induced} by {Valley}
  {Polarization} in {Twisted} {Bilayer} {Graphene}},\ }\href
  {https://doi.org/10.1103/PhysRevLett.130.266003} {\bibfield  {journal}
  {\bibinfo  {journal} {Phys. Rev. Lett.}\ }\textbf {\bibinfo {volume} {130}},\
  \bibinfo {pages} {266003} (\bibinfo {year} {2023})}\BibitemShut {NoStop}%
\bibitem [{\citenamefont {Steiner}\ \emph {et~al.}(2023)\citenamefont
  {Steiner}, \citenamefont {Melischek}, \citenamefont {Trahms}, \citenamefont
  {Franke},\ and\ \citenamefont {von Oppen}}]{Steiner_PRL_2023}%
  \BibitemOpen
  \bibfield  {author} {\bibinfo {author} {\bibfnamefont {J.~F.}\ \bibnamefont
  {Steiner}}, \bibinfo {author} {\bibfnamefont {L.}~\bibnamefont {Melischek}},
  \bibinfo {author} {\bibfnamefont {M.}~\bibnamefont {Trahms}}, \bibinfo
  {author} {\bibfnamefont {K.~J.}\ \bibnamefont {Franke}},\ and\ \bibinfo
  {author} {\bibfnamefont {F.}~\bibnamefont {von Oppen}},\ }\bibfield  {title}
  {\bibinfo {title} {Diode effects in current-biased josephson junctions},\
  }\href {https://doi.org/10.1103/PhysRevLett.130.177002} {\bibfield  {journal}
  {\bibinfo  {journal} {Phys. Rev. Lett.}\ }\textbf {\bibinfo {volume} {130}},\
  \bibinfo {pages} {177002} (\bibinfo {year} {2023})}\BibitemShut {NoStop}%
\bibitem [{\citenamefont {Lu}\ \emph {et~al.}(2023)\citenamefont {Lu},
  \citenamefont {Ikegaya}, \citenamefont {Burset}, \citenamefont {Tanaka},\
  and\ \citenamefont {Nagaosa}}]{Lu_PRL_2023}%
  \BibitemOpen
  \bibfield  {author} {\bibinfo {author} {\bibfnamefont {B.}~\bibnamefont
  {Lu}}, \bibinfo {author} {\bibfnamefont {S.}~\bibnamefont {Ikegaya}},
  \bibinfo {author} {\bibfnamefont {P.}~\bibnamefont {Burset}}, \bibinfo
  {author} {\bibfnamefont {Y.}~\bibnamefont {Tanaka}},\ and\ \bibinfo {author}
  {\bibfnamefont {N.}~\bibnamefont {Nagaosa}},\ }\bibfield  {title} {\bibinfo
  {title} {Tunable josephson diode effect on the surface of topological
  insulators},\ }\href {https://doi.org/10.1103/PhysRevLett.131.096001}
  {\bibfield  {journal} {\bibinfo  {journal} {Phys. Rev. Lett.}\ }\textbf
  {\bibinfo {volume} {131}},\ \bibinfo {pages} {096001} (\bibinfo {year}
  {2023})}\BibitemShut {NoStop}%
\bibitem [{\citenamefont {Banerjee}\ and\ \citenamefont
  {Scheurer}(2024{\natexlab{a}})}]{Banerjee_PRB_2024}%
  \BibitemOpen
  \bibfield  {author} {\bibinfo {author} {\bibfnamefont {S.}~\bibnamefont
  {Banerjee}}\ and\ \bibinfo {author} {\bibfnamefont {M.~S.}\ \bibnamefont
  {Scheurer}},\ }\bibfield  {title} {\bibinfo {title} {Altermagnetic
  superconducting diode effect},\ }\href
  {https://doi.org/10.1103/PhysRevB.110.024503} {\bibfield  {journal} {\bibinfo
   {journal} {Phys. Rev. B}\ }\textbf {\bibinfo {volume} {110}},\ \bibinfo
  {pages} {024503} (\bibinfo {year} {2024}{\natexlab{a}})}\BibitemShut
  {NoStop}%
\bibitem [{\citenamefont {Banerjee}\ and\ \citenamefont
  {Scheurer}(2024{\natexlab{b}})}]{Banerjee_PRL_2024}%
  \BibitemOpen
  \bibfield  {author} {\bibinfo {author} {\bibfnamefont {S.}~\bibnamefont
  {Banerjee}}\ and\ \bibinfo {author} {\bibfnamefont {M.~S.}\ \bibnamefont
  {Scheurer}},\ }\bibfield  {title} {\bibinfo {title} {Enhanced superconducting
  diode effect due to coexisting phases},\ }\href
  {https://doi.org/10.1103/PhysRevLett.132.046003} {\bibfield  {journal}
  {\bibinfo  {journal} {Phys. Rev. Lett.}\ }\textbf {\bibinfo {volume} {132}},\
  \bibinfo {pages} {046003} (\bibinfo {year} {2024}{\natexlab{b}})}\BibitemShut
  {NoStop}%
\bibitem [{\citenamefont {Virtanen}\ and\ \citenamefont
  {Heikkilae}(2024)}]{Virtanen_PRL_2024}%
  \BibitemOpen
  \bibfield  {author} {\bibinfo {author} {\bibfnamefont {P.}~\bibnamefont
  {Virtanen}}\ and\ \bibinfo {author} {\bibfnamefont {T.~T.}\ \bibnamefont
  {Heikkilae}},\ }\bibfield  {title} {\bibinfo {title} {Nonreciprocal josephson
  linear response},\ }\href {https://doi.org/10.1103/PhysRevLett.132.046002}
  {\bibfield  {journal} {\bibinfo  {journal} {Phys. Rev. Lett.}\ }\textbf
  {\bibinfo {volume} {132}},\ \bibinfo {pages} {046002} (\bibinfo {year}
  {2024})}\BibitemShut {NoStop}%
\bibitem [{\citenamefont {Mao}\ \emph {et~al.}(2024)\citenamefont {Mao},
  \citenamefont {Yan}, \citenamefont {Zhuang},\ and\ \citenamefont
  {Sun}}]{Mao_PRL_2024}%
  \BibitemOpen
  \bibfield  {author} {\bibinfo {author} {\bibfnamefont {Y.}~\bibnamefont
  {Mao}}, \bibinfo {author} {\bibfnamefont {Q.}~\bibnamefont {Yan}}, \bibinfo
  {author} {\bibfnamefont {Y.~C.}\ \bibnamefont {Zhuang}},\ and\ \bibinfo
  {author} {\bibfnamefont {Q.-F.}\ \bibnamefont {Sun}},\ }\bibfield  {title}
  {\bibinfo {title} {Universal spin superconducting diode effect from
  spin-orbit coupling},\ }\href
  {https://doi.org/10.1103/PhysRevLett.132.216001} {\bibfield  {journal}
  {\bibinfo  {journal} {Phys. Rev. Lett.}\ }\textbf {\bibinfo {volume} {132}},\
  \bibinfo {pages} {216001} (\bibinfo {year} {2024})}\BibitemShut {NoStop}%
\bibitem [{\citenamefont {Zeng}(2025)}]{Zeng_PRL_2025}%
  \BibitemOpen
  \bibfield  {author} {\bibinfo {author} {\bibfnamefont {W.}~\bibnamefont
  {Zeng}},\ }\bibfield  {title} {\bibinfo {title} {Transverse josephson diode
  effect in tilted dirac systems},\ }\href
  {https://doi.org/10.1103/PhysRevLett.134.176002} {\bibfield  {journal}
  {\bibinfo  {journal} {Phys. Rev. Lett.}\ }\textbf {\bibinfo {volume} {134}},\
  \bibinfo {pages} {176002} (\bibinfo {year} {2025})}\BibitemShut {NoStop}%
\bibitem [{\citenamefont {Chakraborty}\ and\ \citenamefont
  {Black-Schaffer}(2025)}]{Chakraborty_PRL_2025}%
  \BibitemOpen
  \bibfield  {author} {\bibinfo {author} {\bibfnamefont {D.}~\bibnamefont
  {Chakraborty}}\ and\ \bibinfo {author} {\bibfnamefont {A.~M.}\ \bibnamefont
  {Black-Schaffer}},\ }\bibfield  {title} {\bibinfo {title} {Perfect
  superconducting diode effect in altermagnets},\ }\href
  {https://doi.org/10.1103/cv8s-tk4c} {\bibfield  {journal} {\bibinfo
  {journal} {Phys. Rev. Lett.}\ }\textbf {\bibinfo {volume} {135}},\ \bibinfo
  {pages} {026001} (\bibinfo {year} {2025})}\BibitemShut {NoStop}%
\bibitem [{\citenamefont {Qi}\ \emph {et~al.}(2025{\natexlab{b}})\citenamefont
  {Qi}, \citenamefont {Lu}, \citenamefont {Liu}, \citenamefont {Chen},\ and\
  \citenamefont {Xie}}]{Qi_PRB_2025}%
  \BibitemOpen
  \bibfield  {author} {\bibinfo {author} {\bibfnamefont {J.}~\bibnamefont
  {Qi}}, \bibinfo {author} {\bibfnamefont {M.}~\bibnamefont {Lu}}, \bibinfo
  {author} {\bibfnamefont {J.}~\bibnamefont {Liu}}, \bibinfo {author}
  {\bibfnamefont {C.-Z.}\ \bibnamefont {Chen}},\ and\ \bibinfo {author}
  {\bibfnamefont {X.~C.}\ \bibnamefont {Xie}},\ }\bibfield  {title} {\bibinfo
  {title} {Non-hermitian superconducting diode effect},\ }\href
  {https://doi.org/10.1103/n51c-17pn} {\bibfield  {journal} {\bibinfo
  {journal} {Phys. Rev. B}\ }\textbf {\bibinfo {volume} {112}},\ \bibinfo
  {pages} {L060502} (\bibinfo {year} {2025}{\natexlab{b}})}\BibitemShut
  {NoStop}%
\bibitem [{\citenamefont {Tinkham}(1996)}]{Tinkham__1996}%
  \BibitemOpen
  \bibfield  {author} {\bibinfo {author} {\bibfnamefont {M.}~\bibnamefont
  {Tinkham}},\ }\href@noop {} {\emph {\bibinfo {title} {Introduction to
  Superconductivity}}}\ (\bibinfo  {publisher} {{McGraw-Hill, Inc.}},\ \bibinfo
  {year} {1996})\BibitemShut {NoStop}%
\end{thebibliography}%
\end{document}